\documentclass[9pt,twocolumn,twoside]{osajnl}

\journal{josab} 

\setboolean{shortarticle}{false}
\usepackage{physics}
\usepackage{amsmath}
\usepackage{lipsum}
\usepackage{color}

\title{Inverse design of functional photonic patches by adjoint optimization coupled to the generalized Mie theory}

\author[1]{Yilin Zhu}
\author[2]{Yuyao Chen}
\author[2]{Sean Gorsky}
\author[2]{Tornike Shubitidze}
\author[1,2,3,*]{Luca Dal Negro}

\affil[1]{Division of Material Science and Engineering, Boston University, 15 Saint Mary's Street, Brookline, Massachusetts 02446, USA}
\affil[2]{Department of Electrical \& Computer Engineering and Photonics Center, Boston University, 8 Saint Mary's Street, Boston, Massachusetts 02215, USA}
\affil[3]{Department of Physics, Boston University, 590 Commonwealth Avenue, Boston, Massachusetts 02215, USA}

\affil[*]{Corresponding author: dalnegro@bu.edu}

\begin{abstract}
We propose a rigorous approach for the inverse design of functional photonic structures by coupling the adjoint optimization method and the two-dimensional generalized Mie theory (2D-GMT) for the multiple scattering problem of finite-size arrays of dielectric nanocylinders optimized to display desired functions. We refer to these functional scattering structures as "photonic patches". We briefly introduce the formalism of 2D-GMT and the critical steps necessary to implement the adjoint optimization algorithm to photonic patches with designed radiation properties. In particular, we showcase several examples of periodic and aperiodic photonic patches with optimal nanocylinder radii and arrangements for radiation shaping, wavefront focusing in the Fresnel zone, and for the enhancement of the local density of states (LDOS) at multiple wavelengths over micron-size areas. Moreover, we systematically compare the performances of periodic and aperiodic patches with different sizes and find that optimized aperiodic Vogel spiral geometries feature significant advantages in achromatic focusing compared to their periodic counterparts. Our results show that adjoint optimization coupled to 2D-GMT is a robust methodology for the inverse design of compact photonic devices that operate in the multiple scattering regime with optimal desired functionalities. Without the need of spatial meshing, our approach provides efficient solutions at strongly reduced computational burden compared to standard numerical optimization techniques and suggests compact device geometries for on-chip photonics and metamaterials technologies.  
\end{abstract}

\setboolean{displaycopyright}{true} 

\begin{document}

\maketitle
Inverse design is an important methodology for the nanophotonics community that enables developing and prototyping novel devices with desired characteristics and functionalities, greatly enriching the photonic design library beyond standard templates \cite{molesky2018inverse}. In a typical inverse design situation, one first defines an objective function for the system and then applies search algorithms to vary the system's design parameters and optimize an objective function value until it reaches a desired threshold. Gradient-based search algorithms are commonly used that iteratively evaluate the gradient of the objective function with respect to the design parameters and then update these parameters using the gradient information \cite{ruder2016overview, molesky2018inverse}. The adjoint optimization method is a rigorous and general approach that has been widely utilized for the inverse design for photonic devices, such as parametrized metasurfaces \cite{sell2017large, callewaert2018inverse, mansouree2021largescale}, on-chip demultiplexer waveguides \cite{piggott2015inverse, su2018inverse} and nonlinear optical switches \cite{hughes2018adjoint}. The calculations of gradients in forward simulations are typically performed by numerical methods, such as finite-element method (FEM) and finite-difference-time-domain (FDTD) \cite{sell2017large, callewaert2018inverse, piggott2015inverse, su2018inverse, hughes2018adjoint}. However, standard numerical methods are computationally expensive as they require spatial meshing \cite{gagnon_JO_2015}. Therefore, if one could obtain the gradients of the desired system's parameters in analytical or semi-analytical closed-forms, then very efficient adjoint optimizations would be achieved based only on one single forward simulation. Recent examples of analytical gradient calculations include the inverse design of metasurfaces using coupled-mode theory (CMT) \cite{zhou2021inverse}, and the optimization of compact optical elements based on spherical nanoparticles using the multi-sphere generalized Mie theory (3D-GMT) approach \cite{zhan2018inverse, zhelyeznyakov2020design}. Latest techniques also leverage the concept of automatic differentiation (AD) used in artifical neural networks for the inverse design of meta-optics \cite{ma2021deep, colburn2021inverse}.

In this paper, we introduce and utilize the adjoint optimization approach coupled to two-dimensional generalized Mie theory (2D-GMT), which rigorously solves Maxwell's equations for 2D geometries of arbitrary arrays of scattering cylinders. Using this powerful tool, we demonstrate the inverse design of "photonic patches", which are finite-size arrays of nanocylinders with positions and radii efficiently optimized to achieve desired functionalities over small-footprint areas. We remark that rigorous simulations of scattering systems based on 2D-GMT enable the design of aperiodic functional photonic devices based on membrane geometries, which have been fabricated resulting in enhanced light-matter interaction \cite{oliver2020cavity,oliver2021cavity}.

Our paper is organized as follows: in the first section we provide an overview of the 2D-GMT formalism that solves the multiple scattering problem for non-overlapping nanocylinders under an excitation wave perpendicular to the axis of the cylinders. In particular, we discuss analytical closed-form expressions for the far-field scattering intensity and the local density of states (LDOS) that enable the efficient implementation of the adjoint optimization algorithm. Detailed results on the analytical calculations of gradient terms are also provided. Using these results, we provide several application examples of designed photonic patches optimized to perform multi-wavelength radiation shaping, near-field focusing, and to enhance the LDOS over small device areas. Our results demonstrate that the inverse design of photonic patches provides complex optical functionalities over significantly reduced areas compared to traditional photonic crystals and enables scalability advantages for the optical integration of novel aperiodic structures \cite{joannopoulos2008photonic, dal2021aperiodic}. 


\section{Overview of two-dimensional generalized Mie theory (2D-GMT)}
\label{section: 2D-GMT derivation}
In this section, we provide a brief overview of the 2D-GMT formalism by introducing the transfer matrix equation, the scattered far-field amplitude and the local density of states (LDOS). Particular emphasis is placed on closed-form analytical results that enable the efficient calculation of the gradient terms. A detailed derivation and implementation of the 2D-GMT for the nanocylinder array can be found in references \cite{gagnon_JO_2015, dalnegro2022waves}. 

\subsection{Derivation of the transfer matrix equation}\label{subsection: transfermatrix}

The essential idea of 2D-GMT is to expand the fields into a sum of cylindrical Bessel and Hankel functions, which form a complete basis in the 2D domain. Using Graf's addition theorem, we then enforce the electromagnetic boundary conditions at the surface of each cylinder and obtain a matrix equation that relates the known expansion coefficients of the excitation source with the unknown expansion coefficients of the internal and scattered fields. Therefore, the solution of the scattering problem is conveniently formulated as a matrix inversion problem for the unknown field expansion coefficients, as detailed below. 
\begin{figure}[ht!]
	\centering
	\includegraphics[width=0.7\linewidth]{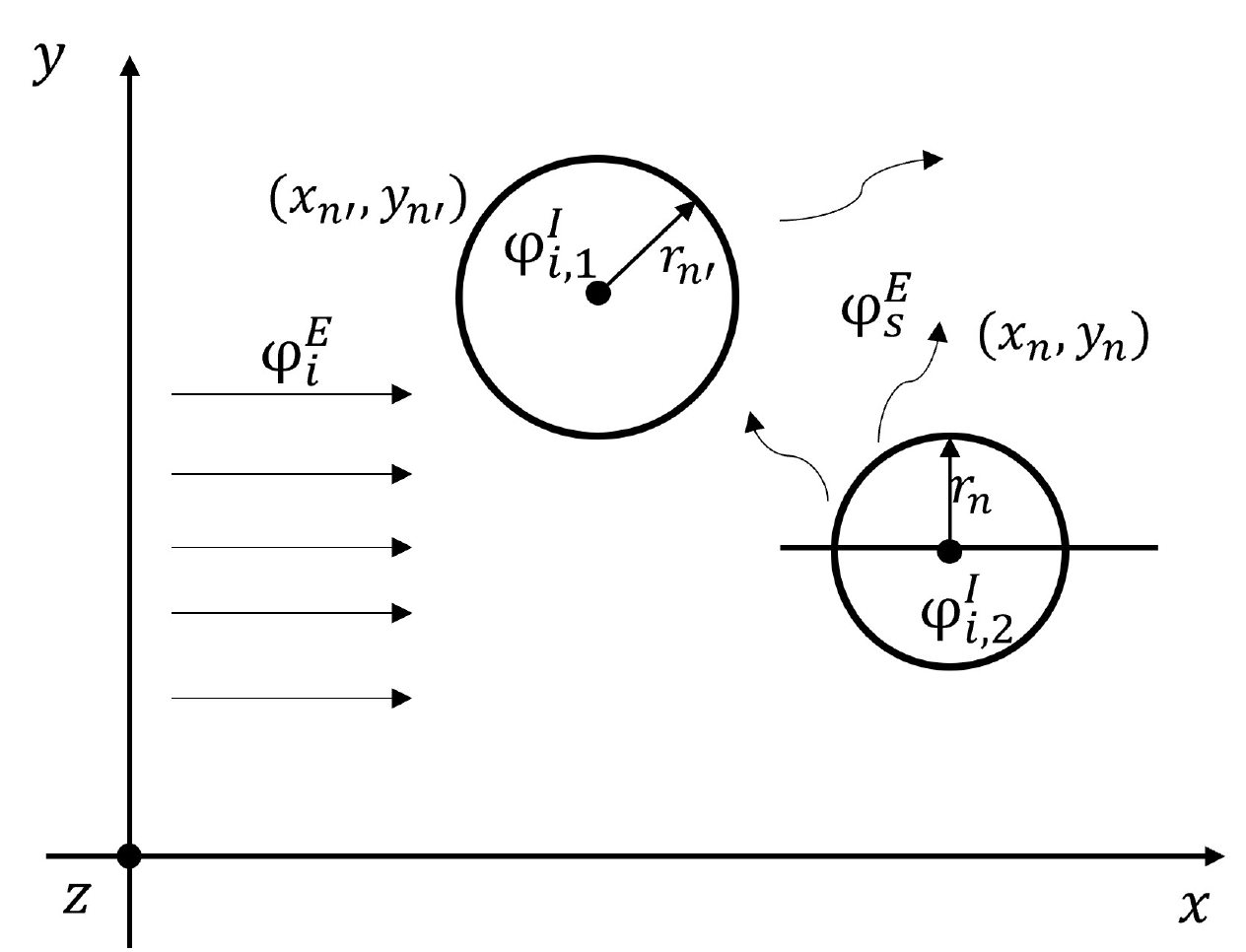}
	\caption{Schematics of the geometry of the scattering problem for two nanocylinders introducing the relevant notation of 2D-GMT. }
	\label{fig:GMT1}
\end{figure}

\begin{figure}[ht!]
	\centering
	\includegraphics[width=\linewidth]{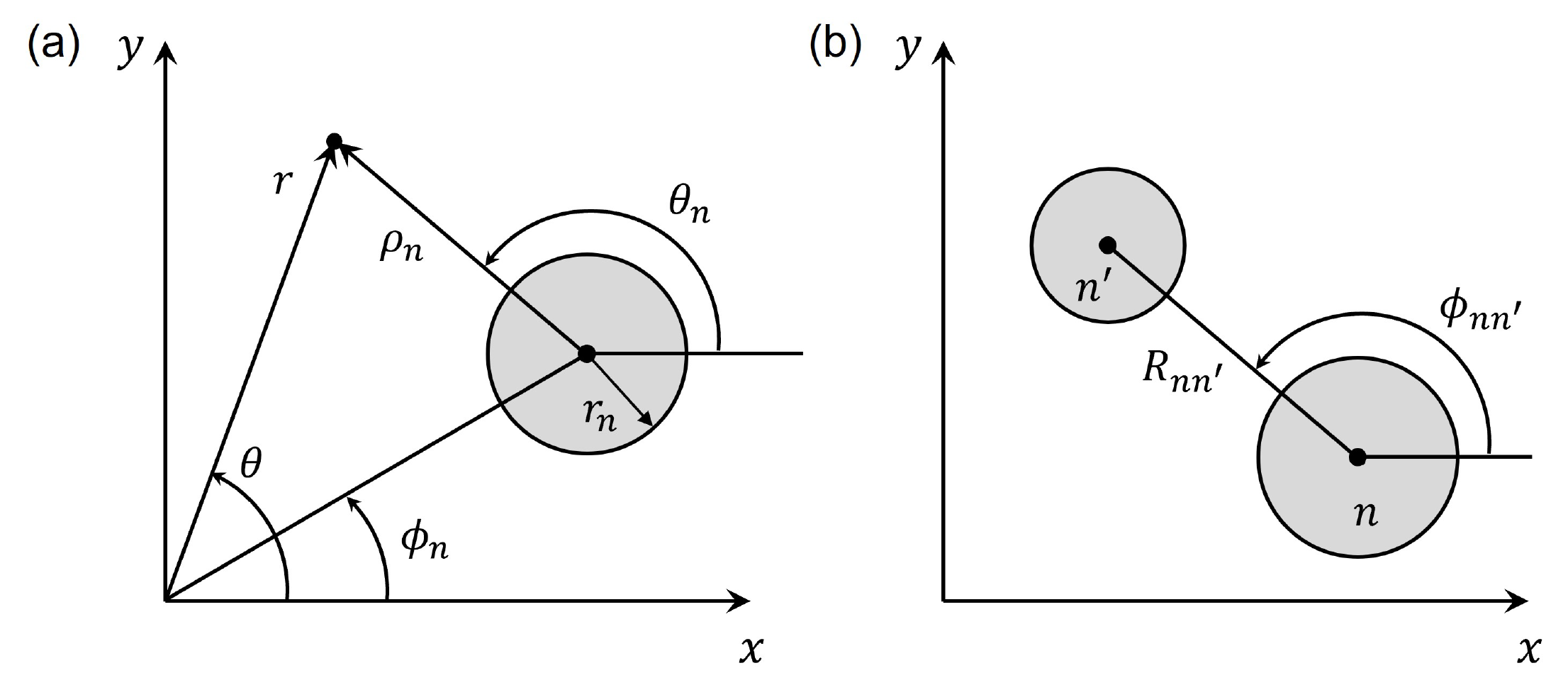}
	\caption{(a) Schematics of the polar coordinates $(r,\theta)$ with origin at $(0,0)$ and polar coordinates $(\rho_n,\theta_n)$ with origin located at the center of the $n\mathrm{th}$ cylinder $(x_n,y_n)$. (b) Illustration of the relation between two local frames used in the derivation of  Eq. \ref{eq:aRelation}.}
	\label{fig:GMT2}
\end{figure}

A typical geometry for which the 2D-GMT solves the scattering problem is displayed in Fig. \ref{fig:GMT1}. In particular, it consists of an aggregate of cylinders with positions $(x_n,y_n)$ and radii $r_n$ with complex relative permittivities $\epsilon_n$, and permeabilities $\mu_n$. The system is embedded in a non-absorbing dielectric host medium with real permittivity $\epsilon_o$ and permeability $\mu_o$. As we restrict the wave propagation to in-plane directions only, the field solutions can be represented as either transverse magnetic (TM) polarization, where the electric field $\mathbf{E}=E_z\hat{\mathbf{z}}$ is oriented along the axis of the cylinders ($z$-axis), or transverse electric (TE) polarization, where $\mathbf{H}=H_z\hat{\mathbf{z}}$ is oriented along $z$-axis. In our discussion, we denote the relevant field component along the cylindrical axis as $\varphi({\mathbf{r}})$, standing either for $E_z$ or for $H_z$, depending on the polarization considered. 
In the schematics shown in Fig. \ref{fig:GMT1}, the exterior field $\varphi^{E}$, which exists only outside nanocylinders, consists of the sum of the incident field $\varphi^{E,\mathrm{inc}}$ and the scattered field $\varphi^{E,\mathrm{sca}}$. These contributions are expanded as an infinite sum of complete basis functions for the cylindrical geometry, which are the  cylindrical Bessel and the Hankel functions. 
Therefore, we represent the exterior field as follows:
\begin{align}
	\varphi^{E}(\mathbf{r}) &= \varphi^{E,\mathrm{sca}}(\mathbf{r}) + \varphi^{E,\mathrm{inc}}(\mathbf{r})
	\label{eq:external}\\
	\varphi^{E,\mathrm{inc}}(\mathbf{r})  &=p_z \sum_{\ell=-\infty}^{\infty}a_{n\ell}^{0E}J_{\ell}(k_{o}\rho_{n})e^{j\ell\theta_{n}} \label{eq:varphiEsrc} \\
	\varphi^{E,\mathrm{sca}}(\mathbf{r}) &=p_z \sum_{n=1}^{N} \sum_{\ell=-\infty}^{\infty} b_{n\ell}H_{\ell}(k_{o}\rho_{n})e^{j\ell\theta_{n}} \label{eq:varphiEsca}
\end{align}    
where $k_o=2\pi\sqrt{\epsilon_o\mu_o}/\lambda$ is the wavenumber in the host medium, $\ell$ is the angular order of the cylindrical functions,  $\mathbf{r}$ is a global position vector, and ($\rho_n$,$\theta_n$) is the local polar coordinate system with its origin located at the center of the $n\mathrm{th}~(n=1,2,\ldots,N)$ cylinder as shown in Fig. \ref{fig:GMT2}(a). The Mie-Lorenz coefficients $a_{n\ell}^{0E}$ depend on the excitation conditions while the coefficients $b_{n\ell}$ are associated to the scattered fields. These quantities are introduced in the local reference frame centered on the $n\mathrm{th}$ nanocylinder. The coefficient $p_z$ is used above to ensure that source properties, such as incident intensity or power, appear to be independent of polarization. Here, we have $p_z=1/Z_0$ for TE polarized excitation and $p_z=1$ for TM polarized excitation, where $Z_0$ is the impedance of the host medium. 

Similarly, the interior field within the $n\mathrm{th}$ nanocylinder $\varphi_n^{I}$ has a contribution originating from the sum of the fields scattered from surfaces of all the other cylinders $\varphi_n^{I,\mathrm{sca}}$, not to be confused with the exterior scattered field $\varphi ^{E,\mathrm{sca}}$, and one from any source that is present inside the $n$th cylinder $\varphi_n^{I,\mathrm{src}}$: 
\begin{align}
	\varphi_n^{I}(\mathbf{r}) &= \varphi_n^{I,\mathrm{src}}(\mathbf{r}) + \varphi_n^{I,\mathrm{sca}}(\mathbf{r})\label{eq: internal}\\
	\varphi_n^{I,\mathrm{src}}(\mathbf{r})  &=p_z \sum_{\ell=-\infty}^{\infty}a_{n\ell}^{0I}H_{\ell}(k_{n}\rho_{n})e^{j\ell\theta_{n}}\\
	\varphi_{n}^{I,\mathrm{sca}}(\mathbf{r}) &=p_{z} \sum_{\ell=-\infty}^{\infty} c_{n \ell} J_{\ell}\left(k_{n} \rho_{n}\right) e^{j \ell \theta_{n}}\label{eq: internalsca}
\end{align}
where $k_n=2\pi\sqrt{\epsilon_n\mu_n}/\lambda$ is the wavenumber inside the $n$th cylinder.  In the expressions above the coefficients $a_{n\ell}^{0I}$ are related to the known source inside the $n\mathrm{th}$ cylinder, if present. 

Based on Eqs. \ref{eq:external} through \ref{eq: internalsca}, the goal of field calculation is to solve for the unknown coefficients $c_{n\ell}$ and $b_{n\ell}$ given the known coefficients $a_{n\ell}^{0E}, a_{n\ell}^{0I}$, through the application of the boundary conditions on the surface of each cylinder. This can be achieved based on the expansion of the exterior field $\varphi^{E}(\mathbf{r})$ in terms of Bessel and Hankel functions centered only on the $n\mathrm{th}$ cylinder:
\begin{equation}
	\label{eq:exterior2}
	\varphi^{E}_n(\mathbf{r})=p_{z} \sum_{\ell}\left[a_{n \ell} J_{\ell}\left(k_{o} \rho_{n}\right)+b_{n \ell} H_{\ell}\left(k_{o} \rho_{n}\right)\right] e^{j \ell \theta_{n}}
\end{equation}

Note that Eq. \ref{eq:exterior2} is derived by applying the Graf's addition theorem that enables the transformation of cylindrical basis functions from the reference frame of the cylinder $n'$ to that of the cylinder $n$ \cite{gagnon_JO_2015,martin2006multiple,dalnegro2022waves}. 
The coefficients $a_{n\ell}$ are expressed as:
\begin{equation} \label{eq:aRelation}
	a_{n \ell}=a_{n \ell}^{0 E}+\sum_{n^{\prime} \neq n} \sum_{\ell^{\prime}=-\infty}^{\infty} e^{j\left(\ell^{\prime}-\ell\right) \phi_{n n^{\prime}}} H_{\ell-\ell^{\prime}}\left(k_{o} R_{n n^{\prime}}\right) b_{n^{\prime} \ell^{\prime}}
\end{equation}
where $(R_{nn'},\phi_{nn'})$ are the polar coordinates of the center of the $n'\mathrm{th}$ cylinder with respect to the frame of reference centered on the $n\mathrm{th}$ cylinder, as shown in Fig. \ref{fig:GMT2}(b). We can now apply the electromagnetic boundary conditions on the surface of each cylinder  $\rho_n=r_n$ according to:
\begin{equation}
	\begin{aligned}
		\varphi_{n}^{I}\left(r_{n}\right) &=\varphi_{n}^{E}\left(r_{n}\right) \\
		\left.\varsigma_{n} \frac{\partial \varphi_{n}^{I}}{\partial \rho_{n}}\right|_{r_{n}}&=\left.\varsigma_{o} \frac{\partial \varphi_{n}^{E}}{\partial \rho_{n}}\right|_{r_{n}}
	\end{aligned}
\end{equation}
where  $\varsigma_i=1/\mu_i$ $[1/\epsilon_i]$ and $\varsigma_o=1/\mu_o$ $[1/\epsilon_o]$ for TM [TE] polarization, respectively. Applying the boundary conditions using Eqs. \ref{eq: internal} and \ref{eq:exterior2} and assuming no internal sources inside the scatterers ($a_{n\ell}^{0I}=0$) we can obtain:
\begin{equation}
	\label{eq:bRelation}
	b_{n\ell}= a_{n\ell}s_{n\ell}
\end{equation}
where:
\begin{equation}
	\begin{aligned}
		s_{n \ell} &=-\frac{J_{\ell}^{\prime}\left(k_{o} r_{n}\right)-\Gamma_{n \ell} J_{\ell}\left(k_{o} r_{n}\right)}{H_{\ell}^{\prime}\left(k_{o} r_{n}\right)-\Gamma_{n \ell} H_{\ell}\left(k_{o} r_{n}\right)} \\
		\Gamma_{n \ell} &=\frac{\xi_{n} k_{n} J_{\ell}^{\prime}\left(k_{n} r_{n}\right)}{k_{o} J_{\ell}\left(k_{n} r_{n}\right)} \\
		\xi_{n} &=\frac{\mu_{o}}{\mu_{n}} ~ \left[\frac{\epsilon_{o}}{\epsilon_{n}}\right] ~ \text { for } \mathrm{TM}~[\mathrm{TE}]
	\end{aligned}
\end{equation}
Here the prime symbol denotes the first derivative of the corresponding function with respect to its entire argument. Substituting Eq. \ref{eq:aRelation} into Eq. \ref{eq:bRelation} yields the relation between $b_{n \ell}$ and $a_{n \ell}^{0 E}$ as follows:
\begin{eqnarray}
	b_{n \ell}-s_{n \ell} \sum_{n^{\prime} \neq n} \sum_{\ell^{\prime}=-\infty}^{\infty}e^{j\left(\ell^{\prime}-\ell\right) \phi_{n n^{\prime}}} &H_{\ell-\ell^{\prime}}\left(k_{o} R_{n n^{\prime}}\right) b_{n^{\prime} \ell^{\prime}}\nonumber\\&=s_{n \ell} a_{n \ell}^{0 E}
\end{eqnarray}
Such a relation can be written in matrix form as:
\begin{equation}
	\mathbf{Tb} =	\boldsymbol{\mathrm{a}}^0
\end{equation}
where we have introduced the transfer matrix or T matrix with elements: \begin{equation}
	\mathbf{T}_{n n^{\prime}}^{\ell \ell^{\prime}}=\delta_{n n^{\prime}} \delta_{\ell \ell^{\prime}}-\left(1-\delta_{n n^{\prime}}\right) e^{j\left(\ell^{\prime}-\ell\right) \phi_{n n^{\prime}}} H_{\ell-\ell^{\prime}}\left(k_{o} R_{n n^{\prime}}\right) s_{n \ell}
\end{equation}
Here $\delta$ is the Kronecker function and we introduced the vector notation:
\begin{equation}
	\mathbf{a}^{0}=\left\{a_{n \ell}\right\}=\left\{s_{n \ell} a_{n \ell}^{0 E}\right\}, \quad \mathbf{b}=\left\{b_{n \ell}\right\} .
\end{equation}

In practical implementations, we must limit the range of $\ell$ in the angular expansions to a specified cutoff order and consider terms ranging from  $-\ell_{\mathrm{max}}$ to $\ell_{\mathrm{max}}$. A larger $\ell_{\mathrm{max}}$ value guarantees a more accurate solution but adds computational cost in the solution the scattering problem. Therefore, in any given situation it is important to identify a suitable $\ell_{\mathrm{max}}$ value through a detailed convergence analysis. Moreover, in order to improve the accuracy of the numerical results, we follow reference \cite{gagnon_JO_2015} and solve for the scaled equations as follows:
\begin{eqnarray} 
	\hat{\mathbf{T}} \hat{\mathbf{b}} =&&\hat{\mathbf{a}}^{0} \label{eq: 2D-GMT matrix equation}\\
	\hat{\mathbf{b}}  = &&b_{n \ell} / J_{\ell}\left(k_{o} r_{n}\right) \\
	\hat{\mathbf{a}}^{0}  = &&a_{n \ell} / J_{\ell}\left(k_{o} r_{n}\right) \\
	\hat{\mathbf{T}} =&& \delta_{n n^{\prime}} \delta_{\ell \ell^{\prime}}-\left(1-\delta_{n n^{\prime}}\right) e^{j\left(\ell^{\prime}-\ell\right) \phi_{n n^{\prime}}} \nonumber \\
	&& \times H_{\ell-\ell^{\prime}}\left(k_{o} R_{n n^{\prime}}\right) s_{n \ell} \frac{J_{\ell^{\prime}}\left(k_{o} r_{n^{\prime}}\right)}{J_{\ell}\left(k_{o} r_{n}\right)}
\end{eqnarray}
Detailed expressions for the incident wave coefficients ${\mathbf{a}}^{0}$ of plane waves, collimated source beams (similar to Gaussian beams but strictly a solution of the 2D Helmholtz equation), and the excitation dipoles can be found in the references \cite{gagnon_JO_2015,asatryan2003}. 

\subsection{Derivation of the scattered far-field amplitude}
\label{subsection: scattered far-field amplitude}
Once the scattered field of the array is obtained by Eq. \ref{eq:varphiEsca}, we can express in closed-form relevant far-field quantities used in the analysis of wave scattering systems. In particular, we focus here on the scattering amplitude $F_z^\mathrm{sca}(\theta)$, which is defined through the asymptotic far-field expression: 
\begin{equation}
	\varphi_{z}^\mathrm{sca}(r,\theta) = F_{z}^\mathrm{sca}(\theta)\frac{e^{jk_{o}r}}{\sqrt{r}}
	\label{eq:amplitudeRelation}
\end{equation}
where $r=|\mathbf{r}|$. The scattering amplitude can be derived from the scattered fields by evaluating Eq. \ref{eq:varphiEsca} in the limit of $r\rightarrow \infty$. Specifically, considering the asymptotic form of the Hankel function $H_{\ell}(z) \sim \sqrt{\frac{2}{\pi z}} e^{j(z-\ell \pi / 2-\pi / 4)}$ we obtain:
\begin{equation}\label{eq: far-field1}
	\varphi_{z}^\mathrm{sca}(\mathbf{r})\approx p_z\sum_{n=1}^{N} \sum_{\ell=-\ell_{\mathrm{max}}}^{\ell_{\mathrm{max}}}\sqrt{\frac{2}{\pi k_{o}\rho_{n}}}e^{j(k_{o}\rho_{n}-\ell\pi/2-\pi/4 + \ell\theta_{n})}
\end{equation}
Moreover, using the cosine law on the triangle shown in Fig. \ref{fig:GMT2}(a) we find that $\rho_{n}=\sqrt{r^{2} + R_{n}^{2}-r R_{n}cos(\theta-\phi_{n})}$. In the far-field limit, $r\rightarrow \infty$, we further obtain:
\begin{equation}
	\rho_{n} \approx r - R_{n}\cos(\theta - \phi_{n})
\end{equation}

Although the $\cos(\theta-\phi_{n})$ term is small relative to $r$ in the far-field, it has a significant impact on the phasor term of Eq. \ref{eq: far-field1}, and thus it must be kept in the exponent. On the other hand, we can directly substitute $\rho_n$ with $r$ in the prefactor of Eq. \ref{eq: far-field1}. Therefore, we obtain:
\begin{eqnarray}
	\varphi_{z}^\mathrm{sca}(\mathbf{r})\approx &&p_z\frac{e^{jk_{o}r}}{\sqrt{r}} \sqrt{\frac{2}{\pi k_{o}}}  \sum_{n=1}^{N}\sum_{\ell=-\ell_{\mathrm{max}}}^{\ell_{\mathrm{max}}} \bigg[ b_{n\ell}\nonumber\\
	&&\times \left. e^{-j[k_{o}R_{n}\cos(\theta-\phi_{n}) +\ell(\frac{\pi}{2}-\theta)+\frac{\pi}{4}]}\right]
	\label{eq:scatteredFarfield}
\end{eqnarray}
Finally, comparing Eqs. \ref{eq:amplitudeRelation} and \ref{eq:scatteredFarfield} yields the expression for the scattering amplitude:
\begin{eqnarray}
	\label{eq:Fzsca}
	F_{z}^{\mathrm{sca}}(\theta)=&&p_z\sqrt{\frac{2}{\pi k_{o}}}\sum_{n=1}^{N}\sum_{\ell=-\ell_{\mathrm{max}}}^{\ell_{\mathrm{max}}}\bigg[ b_{n\ell}\nonumber\\
	&&\times \left. e^{-j[k_{o}R_{n}\cos(\theta-\phi_{n}) +\ell(\frac{\pi}{2}-\theta)+\frac{\pi}{4}]}\right]
\end{eqnarray}

Based on Eq. \ref{eq:Fzsca} we can further derive a closed-form analytical expression for the scattered far-field angular intensity, which is a key quantity of interest in directional radiation problems. The far-field angular intensity is found by substituting Eqs. \ref{eq:amplitudeRelation} and \ref{eq:Fzsca} into the time-averaged Poynting vector expression: 
\begin{equation}\label{eq:Ssca}
	\langle\mathbf{S}^\mathrm{sca}\rangle=\frac{1}{2}\mathrm{Re}\left[\mathbf{E}^\mathrm{sca}\times\mathbf{H}^\mathrm{sca}\right]
\end{equation}
To remove the radial dependence of the far-field intensity, we multiply Eq. \ref{eq:Ssca} by the radial distance $r$. The far-field angular intensity is thus given by the limit of the product when $r\rightarrow\infty$, resulting in: 
\begin{equation}
	I^\mathrm{sca}(\theta)=\lim_{r\rightarrow\infty}r\langle\mathbf{S}^\mathrm{sca}\rangle=\frac{1}{2Z_o p_z^2}|F_z^{\mathrm{sca}}(\theta)|^2
\end{equation}

Based on the knowledge of the far-field angular intensity, one can obtain quantitative information on the directional scattering properties of the arrays  through their differential scattering cross-section. This quantity is obtained by normalizing the far-field angular intensity by the incident intensity $I_o$. In 2D-GMT calculation $I_o=1/(2Z_o)$ and we obtain the following expression for the differential scattering cross section:
\begin{equation}
	\frac{\partial \sigma^\mathrm{sca}}{\partial \theta}=\frac{I^\mathrm{sca}(\theta)}{I_o}=\frac{|F^\mathrm{sca}_z|^2}{p_z^2}\label{eq: differsca}
\end{equation}
The differential scattering cross section describes how efficiently the incident radiation is scattered along a given angular direction and it is of great importance in evaluating the performances of devices used for radiation engineering \cite{forouzmand2018tunable}. 

\subsection{Derivation of the local density of states}

The local density of states (LDOS) quantifies the number of electromagnetic modes into which photons of a given wavelength can be emitted at a specified position in space. The LDOS is particularly useful because it is related to experimentally observable quantities such as transmission gaps and the spontaneous decay rate of embedded light sources inside non-homogeneous photonic environments \cite{sprik1996,novotny2012principles,dal2021aperiodic}. Moreover, by comparing the LDOS in a photonic device to the one in free-space we can characterize the degree of enhancement or suppression of light emission. 

The LDOS is related to the imaginary part of the electric field Green tensor \cite{balian1971}:
\begin{equation}\label{eq:LDOS}
    \rho(\mathbf{r};\lambda)=-\frac{4 n_o^2}{c\lambda}\Im\{  \textrm{Tr}[\boldsymbol{\mathrm{G}}^{e}(\mathbf{r},\mathbf{r};\lambda)]\}
\end{equation}
where $n_o=\sqrt{\epsilon_{o}\mu_{o}}$ is the refractive index of host medium, $\Im\{\cdot\}$ denotes the imaginary part of a complex quantity and $\mathrm{Tr}$ denotes the trace operation. The electric field Green tensor $\boldsymbol{\mathrm{G}}^{e}(\mathbf{r},\mathbf{r}_s;\lambda)$ is the electric field response at spatial location $\mathbf{r}=(x,y)$ resulting from a source at position $\mathbf{r}_s=(x_s,y_s)$. In general, it is a second-rank tensor where the elements in column $u$ represent the components of the total electric field vector $(G_{xu}, G_{yu}, G_{zu})^\mathrm{T}$ generated by the dipole source with orientation parallel to the $u=x,y,z$ axes \cite{asatryan2003}.

In the 2D-GMT formalism, the trace of the Green tensor equals the total electric field component located at the source position $(x_s,y_s)$, along the given dipole orientation. Therefore, depending on the polarization of the dipole source, the LDOS can be written as \cite{asatryan2003}:
\begin{eqnarray}
	\rho_\mathrm{TE}=&& -\frac{4 n_o^2}{c\lambda}\Im\{  G_{xx}+G_{yy} \} \label{eq: rhoTE}\\ 
	\rho_\mathrm{TM}=&& -\frac{4 n_o^2}{c\lambda}\Im\{  G_{zz} \} \label{eq: rhoTM} 
\end{eqnarray}
We emphasize here that the total electric field is equal to the sum of the scattered field and the incident field generated by the dipole source, which is:
\begin{equation}
	G_{uu}=E^\mathrm{inc}_u+E^\mathrm{sca}_u\quad u=x,y,z\label{eq: Guu}
\end{equation}

We have derived the $z$-component of the exterior scattered electric field in Eq. \ref{eq:varphiEsca}. The corresponding $x$- and $y$-component of the exterior scattered field can be readily obtained from the dynamic Maxwell's equations: 
\begin{equation}
	-j\omega\epsilon\mathbf{E}^\mathrm{sca}=\nabla\times\mathbf{H}^\mathrm{sca}=\frac{\partial H_z^\mathrm{sca}}{\partial y}\hat{\mathbf{x}}-\frac{\partial H_z^\mathrm{sca}}{\partial x}\hat{\mathbf{y}}\label{eq: Etransverse}
\end{equation}
On the other hand, the components of the field excited by a 2D dipole (i.e., a line source) at any position have been obtained in reference \cite{asatryan2003} as: 
\begin{eqnarray}
	E^\mathrm{inc}_x= && -\frac{j}{8}\left[H_0(k_o\rho)+H_2(k_o\rho)\cos(2\theta)\right] \label{Eq:Eincx}\\ 
	E^\mathrm{inc}_y =&& -\frac{j}{8}\left[H_0(k_o\rho)-H_2(k_o\rho)\cos(2\theta)\right] \label{Eq:Eincy} \\
	E^\mathrm{inc}_z=&& -\frac{j}{4}H_0(k_o\rho) \label{Eq:Eincz}
\end{eqnarray}
where $(\rho,\theta)$ are polar coordinates centered at source location $(x_s,y_s)$. Applying Graf's theorem to Eqs. \ref{Eq:Eincx} through \ref{Eq:Eincz}, we can obtain the source coefficients of the dipole: 
\begin{eqnarray}
	a_{n\ell,x}^{0E}= && -\frac{1}{8j}\left[H_{\ell+1}(k_oR_{n s})e^{-j (\ell+1) \theta_{n s}}\right. \nonumber \\
	&& + \left. H_{\ell-1}(k_oR_{n s})e^{-j (\ell-1) \theta_{n s}} \right] \label{Eq:a0Ex}\\ 
	a_{n\ell,y}^{0E} =&& -\frac{1}{8j}\left[H_{\ell+1}(k_oR_{n s})e^{-j (\ell+1) \theta_{n s}}\right. \nonumber \\
	&& - \left. H_{\ell-1}(k_oR_{n s})e^{-j (\ell-1) \theta_{n s}} \right] \label{Eq:a0Ey}\\ 
	a_{n\ell,z}^{0E}= && \frac{1}{4j}H_0(k_oR_{n s})e^{-j \ell \theta_{n s}} \label{Eq:a0Ez}
\end{eqnarray}
where $R_{ns}=\sqrt{(x_s-x_n)^2+(y_s-y_n)^2}$ and $\theta_{ns}=\tan^{-1}(\frac{y_s-y_n}{x_s-x_n})$. 

Substituting Eqs. \ref{eq: Etransverse} through \ref{Eq:a0Ez} into Eqs. \ref{eq: Guu}, we obtain the expressions for the total field at the excitation dipole position with different orientations:
\begin{eqnarray}
    G_{x x}= &&-\frac{j}{8} +\sum_{n \ell} j b_{n \ell} e^{j \ell \theta_{n s}}\bigg[H_{\ell}^{\prime}\left(k_{o} R_{n s}\right) \sin \left(\theta_{n s}\right) \nonumber \\
    && + \left.\frac{j \ell}{k_{o} R_{n s}} H_{\ell}\left(k_{o} R_{n s}\right) \cos \left(\theta_{n s}\right)\right] \label{Eq:Gxx}\\ 
    G_{y y}=&& -\frac{j}{8} -\sum_{n \ell} j b_{n \ell} e^{j \ell \theta_{n s}}\bigg[H_{\ell}^{\prime}\left(k_{o} R_{n s}\right) \cos \left(\theta_{n s}\right)\nonumber \\ && \left. -\frac{j \ell}{k_{o} R_{n s}} H_{\ell}\left(k_{o} R_{n s}\right) \sin \left(\theta_{n s}\right)\right] \label{Eq:Gyy} \\
    G_{z z}=&& -\frac{j}{4} +\sum_{n \ell} b_{n \ell} H_{\ell}\left(k_{o} R_{n s}\right) e^{j \ell \theta_{n s}} \label{Eq:Gzz}
\end{eqnarray}
where $\sum_{n \ell} \equiv \sum_{n=1}^{N}\sum_{\ell=-\ell_{\mathrm{max}}}^{\ell_{\mathrm{max}}}$. Note that the constant terms in the equations above originate from selecting the observation point exactly at the source location, i.e., by setting $\rho=0$ in Eqs. \ref{Eq:Eincx} through \ref{Eq:Eincz}. 

Based on the expressions above for the LDOS, we can obtain the Purcell enhancement factor $\mathrm{F}(\mathbf{r};\lambda)$, which characterizes the modification of the LDOS in the presence of a structured photonic environment with respect to a homogeneous medium, here assumed to be free space. The Purcell factor is generally defined as:
\begin{equation}\label{eq:purcellEnhancement}
    \mathrm{F}(\mathbf{r};\lambda)=\frac{\rho(\mathbf{r};\lambda)}{\rho_0(\mathbf{r};\lambda)}=-4\: \Im \left\{\mathrm{Tr} [\boldsymbol{\mathrm{G}}^{e}(\mathbf{r},\mathbf{r};\lambda)]\right\}=\frac{\Gamma(\mathbf{r};\lambda)}{\Gamma_0(\mathbf{r};\lambda)}
\end{equation}
where $\rho_0$ is the LDOS of the homogeneous host medium, $\Gamma_0$ is the decay rate of a dipole in the homogeneous medium, and $\Gamma$ is its decay rate in the structured environment. Notice that $\mathrm{F}(\mathbf{r};\lambda)$ becomes unity when the source is embedded in the homogeneous medium. On the other hand, $\mathrm{F}(\mathbf{r};\lambda)>1$ indicates that the photonic structure enhances the radiative properties of the dipole, while spontaneous emission is suppressed when $\mathrm{F}(\mathbf{r};\lambda)<1$.

\begin{figure}[ht!]
	\centering
	\includegraphics[width=\linewidth]{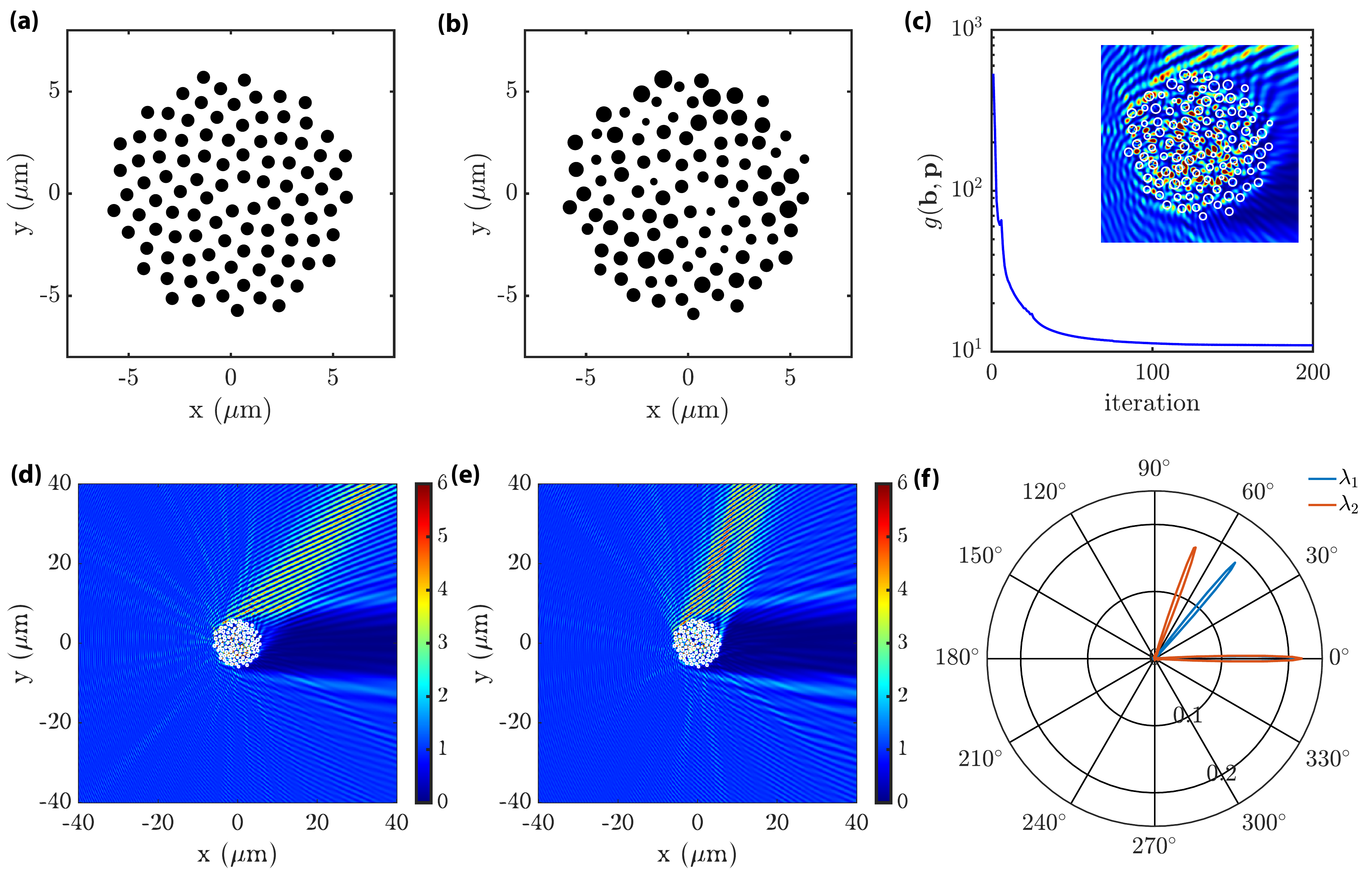}
	\caption{Example of a photonic patch consisting of an optimized GA Vogel spiral for steering two incident wavelengths into two directions. (a) Initial geometry of the GA Vogel spiral structure. (b) Optimized geometry of the photonic patch. (c) Objective function value as a function of the number of iterations. The inset shows the total electric field inside the photonic patch. In panels (d) and (e) we show the total electric field intensity distribution under TM plane wave excitation at wavelengths $\lambda_1$ and $\lambda_2$, respectively. (f) Polar plot of the scattered far-field intensity.}
	\label{fig:scattering_VS}
\end{figure}

\section{Adjoint optimization coupled with 2D-GMT}
\label{section: adjoint 2D-GMT}

In this section, we provide an overview of the general adjoint optimization method and discuss the details of its coupling to the 2D-GMT when applied to finite-size arrays of dielectric nanocylinders. We will then show how to inverse design photonic patches engineered to shape the far-field radiation, focus incident radiation in the Fresnel zone, and enhance the LDOS and the quality factor $Q$ of resonant modes at different wavelengths. 

\subsection{The adjoint optimization method}
\label{subsection: adjoint 2D-GMT}
Suppose we define an objective function $\mathrm{g}(\hat{\mathbf{b}}, \mathbf{p})$ that depends on both the scattered field coefficients $\hat{\mathbf{b}}$ and a vector of design parameters $\mathbf{p}$. These may include the positions and radii of each cylinder, their composition, etc. The key quantity to compute is the gradient of $\mathrm{g}$ with respect to $\mathbf{p}$, which we write as:
\begin{equation}
    \label{eq:delg}
    \nabla_\mathbf{p} \mathrm{g}(\hat{\mathbf{b}}, \mathbf{p})=\mathrm{g}_{\mathbf{p}}+\mathrm{g}  _{\hat{\mathbf{b}}} \hat{\mathbf{b}}_{\mathbf{p}}
\end{equation}
where the subscript symbols indicate partial derivative operations with respect to those quantities, i.e., $\mathrm{g}_{\mathbf{p}}= {\partial\mathrm{g}}/{\partial\mathbf{p}}$, $\mathrm{g}_{\hat{\mathbf{b}}}= \partial\mathrm{g}/\partial{\hat{\mathbf{b}}}$, and $\hat{\mathbf{b}}_{\mathbf{p}}= \partial\hat{\mathbf{b}}/\partial{\mathbf{p}}$. Notice that the term $\hat{\mathbf{b}}_{\mathbf{p}}$ is generally computationally expensive to evaluate when using any full numerical method, such as the finite difference or the finite element method, as it requires $N\gg1$ simulations for each design parameter stored in the vector $\mathbf{p}$ \cite{molesky2018inverse}. However, as we have shown in Section \ref{section: 2D-GMT derivation}, the 2D-GMT solves the scattering problem analytically with the T matrix equation Eq. \ref{eq: 2D-GMT matrix equation}, and provides efficient evaluation of closed-form solutions for the forward simulations. In order to leverage this advantage, we first take the derivative with respect to $\mathbf{p}$ on both sides of Eq. \ref{eq: 2D-GMT matrix equation}, which yields:
\begin{equation}
    \hat{\mathbf{T}}_{\mathbf{p}} \hat{\mathbf{b}}+\hat{\mathbf{T}} \hat{\mathbf{b}}_{\mathbf{p}}=\hat{\mathbf{a}}_{\mathbf{p}}^{0}
\end{equation}
where $\hat{\mathbf{T}}_{\mathbf{p}} = \partial\hat{\mathbf{T}}/\partial{\mathbf{p}}$ and $\hat{\mathbf{a}}_{\mathbf{p}}^{0}= \partial\hat{\mathbf{a}}^{0}/\partial{\mathbf{p}}$. After rearranging the terms, we obtain:
\begin{equation} \label{eq: adjoint1}
    \hat{\mathbf{b}}_{\mathbf{p}}=\hat{\mathbf{T}}^{-1}\left[\hat{\mathbf{a}}_{\mathbf{p}}^{0}-\hat{\mathbf{T}}_{\mathbf{p}} \hat{\mathbf{b}}\right]
\end{equation}
Crucially, substituting Eq. \ref{eq: adjoint1} in Eq. \ref{eq:delg}, we can write the expression for the gradient term:
\begin{equation}
    \nabla_\mathbf{p} \mathrm{g}(\hat{\mathbf{b}}, \mathbf{p})=\mathrm{g}_{\mathbf{p}}+\mathrm{g}_{\hat{\mathbf{b}}}\left(\hat{\mathbf{T}}^{-1}\left[\hat{\mathbf{a}}_{\mathbf{p}}^{0}-\hat{\mathbf{T}}_{\mathbf{p}} \hat{\mathbf{b}}\right]\right)
\end{equation}
The equation above, which enables the efficient calculation of the parameterized gradient within the T matrix formalism, is the main result of this section. This result is often expressed in the literature as:
\begin{align}
    \nabla_\mathbf{p} \mathrm{g} &=\mathrm{g}_{\mathbf{p}}+\left(\boldsymbol{\lambda}^{\mathrm{T}}\left[\hat{\mathbf{a}}_{\mathbf{p}}^{0}-\hat{\mathbf{T}}_{\mathbf{p}} \hat{\mathbf{b}}\right]\right) \label{eq:delgadjoint} \\
    \hat{\mathbf{T}}^{\mathrm{T}} \boldsymbol{\lambda} &=\mathrm{g}_{\hat{\mathbf{b}}}^{\mathrm{T}}\label{eq:adjoint}
\end{align}
where $\boldsymbol{\lambda}=\left(\hat{\mathbf{T}}^{\mathrm{T}}\right)^{-1}\mathrm{g}_{\hat{\mathbf{b}}}^{\mathrm{T}}$, the superscript $\text{T}$ indicates the transpose operation, and the Eq. \ref{eq:adjoint} is referred to as the adjoint equation \cite{molesky2018inverse}. Eqs. \ref{eq:delgadjoint} and \ref{eq:adjoint} enable the efficient computation of the gradient based on only a single forward simulation to obtain the coefficients $\mathbf{\hat{b}}$. The derivative quantities $\hat{\mathbf{T}}_{\mathbf{p}}, \hat{\mathbf{a}}_{\mathbf{p}}^{0}$ are evaluated analytically using the previously established results of the 2D-GMT theory and their explicit expressions are provided in Appendix \ref{appendix: T derivative} through \ref{appendix: dipole derivative}. In the next subsection, we will discuss the explicit calculations of  the derivatives $\mathrm{g}_{\hat{\mathbf{b}}}$ and $\mathrm{g}_{\mathbf{p}}$ for different choices of the objective function associated to different properties of interest of the cylinder arrays and we will also discuss specific optimization cases. 

\subsection{Inverse design of photonic patches for radiation shaping}
\label{subsection: AO radiation shaping}
In this subsection, we apply the adjoint method to the specific design of "photonic patches", which are compact arrays of nanocylinders ($\sim100$ elements) that occupy a small footprint area and exhibit an optimal functionality. In particular, we begin by presenting our results on the design of photonic patches that can efficiently steer incoming radiation of different wavelengths into desired far-field angles. 

\begin{figure}[ht!]
	\centering
	\includegraphics[width=\linewidth]{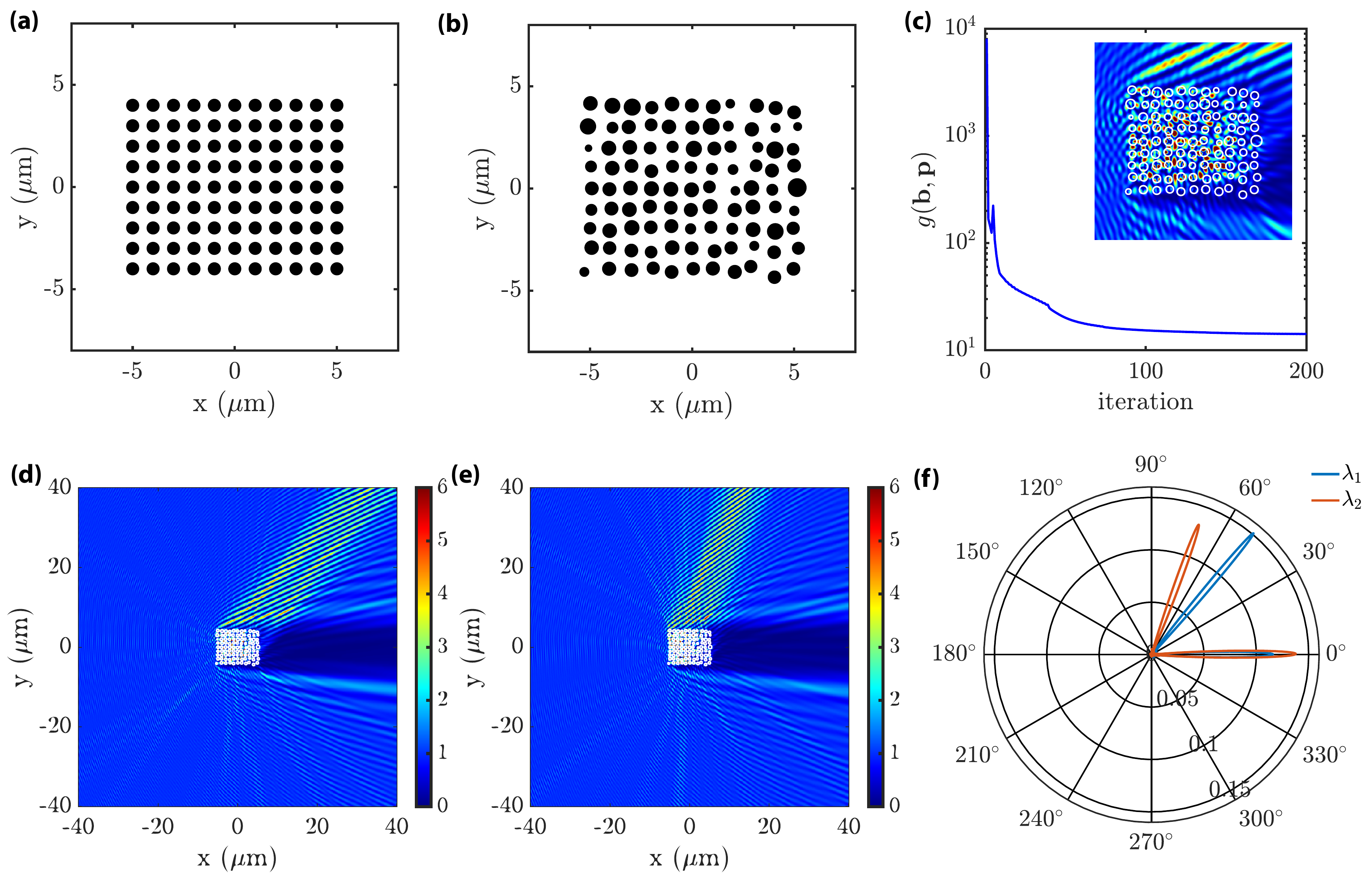}
	\caption{Example of a photonic patch optimized starting from a periodic geometry that steers two wavelengths into two directions. (a) Initial geometry of the periodic structure. (b) Optimized photonic patch geometry. (c) Objective function value as a function of the number of iterations. The inset shows the total electric field inside the photonic patch. In panels (d) and (e) we show the total electric field intensity distribution under TM plane wave excitation at wavelengths $\lambda_1$ and $\lambda_2$, respectively. (f) Polar plot of the scattered far-field intensity.}
	\label{fig:scattering_periodic}
\end{figure}

In order to optimize the directional radiation properties of photonic patches we consider the objective function defined by the scattering intensity at the desired angle $\theta_o$ and wavelength $\lambda_o$:
\begin{equation}
    \label{eq:gIsca}
    \mathrm{g}(\mathbf{r},\mathbf{\hat{b}})=I^{\mathrm{sca}}(\theta_o,\lambda_o)=\frac{1}{2Z_o p_z^2}|F_z^{\mathrm{sca}}(\theta_o,\lambda_o)|^2
\end{equation}
where $F_z^{\mathrm{sca}}(\theta)$ is the scattering amplitude provided in Eq. \ref{eq:Fzsca}. In order to apply the adjoint optimization method, we first need to obtain the expressions for $\mathrm{g}_{\mathbf{p}}$ and $\mathrm{g}_{\hat{\mathbf{b}}}$. For the $i\mathrm{th}~(i=1,2,\ldots,P)$ component of the vector $\mathrm{g}_{\mathbf{p}}$, we get:
\begin{equation}
    \frac{\partial I(\theta_o,\lambda_o)}{\partial p_{i}}=\frac{1}{Z_{o} p_{z}^{2}} \Re\left\{\left(F_{z}^{\mathrm{sca}}\right)^{*} \frac{\partial F_{z}^{\mathrm{sca}}}{\partial p_{i}}\right\}
\end{equation}
where $\Re\{\cdot\}$ denotes the real part of the complex quantity and $^*$ denotes its complex conjugate. Combining this expression with Eq. \ref{eq:Fzsca}, we obtain:
\begin{eqnarray}
    &&\frac{\partial I(\theta)^{\mathrm{sca}}}{\partial p_{i}} =\frac{-j k_{o}}{Z_{o}}\left(\sum_{n \ell} \hat{b}_{n \ell} \gamma_{n \ell}\right)^{*} \times\left(\sum_{n \ell} \hat{b}_{n \ell} \gamma_{n \ell}\left[k_{o} \frac{J_{\ell}^{\prime}\left(k_{o} r_{n}\right)}{J_{\ell}\left(k_{o} r_{n}\right)}\right.\right. \nonumber\\
    && \times \frac{\partial r_{n}}{\partial p_{i}} \left.\left.+R_{n} \sin \left(\theta-\phi_{n}\right) \frac{\partial \phi_{n}}{\partial p_{i}}+\cos \left(\theta-\phi_{n}\right) \frac{\partial R_{n}}{\partial p_{i}}\right]\right)
\end{eqnarray}
where $\gamma_{n \ell}=\sqrt{\frac{2}{\pi k_{o}}} J_{\ell}\left(k_{o} r_{n}\right) e^{-j\left[k_{o} R_{n} \cos \left(\theta-\phi_{n}\right)+\ell(\pi / 2-\theta)+\pi / 4\right]}$. The expressions for the derivatives of the geometrical parameters of the array with respect to the considered design parameters, i.e., the positions and radii of each cylinder, can be found in Table \ref{table:derivativesCylGeom}.

\begin{table}[h!]
    \caption{\label{table:derivativesCylGeom}%
        Derivatives of the array geometry with respect to the design parameters.}
    \begin{tabular}{lll}
        \hline
        ${\partial r_{n}}/{\partial x_j} = 0$ &
        ${\partial r_n}/{\partial y_j} =0$ &
        ${\partial r_n}/{\partial r_j} =\delta_{ni}$\\
        ${\partial R_{n}}/{\partial x_j} = \cos(\phi_n)\delta_{nj}$ &
        ${\partial R_n}/{\partial y_j} =\sin(\phi_n)\delta_{nj}$ &
        ${\partial R_n}/{\partial r_j} = 0$\\
        $\frac{\partial \phi_{n}}{\partial x_j} = -\frac{\sin(\phi_n)}{R_n}\delta_{nj}$ &
        $\frac{\partial \phi_n}{\partial y_j} = \frac{\cos(\phi_n)}{R_n}\delta_{nj}$ &
        $\frac{\partial \phi_n}{\partial r_j} =0$\\
        \hline
    \end{tabular}
\end{table}
In Table \ref{table:derivativesCylGeom}, the design parameters $(x_j,y_j)$ and $r_j$ correspond to the center coordinates and radius of each $j\mathrm{th}$ cylinder in the array. The expression for $\mathrm{g}_{\hat{\mathbf{b}}}$ can be obtained from Eq. \ref{eq:Fzsca} as follows:
\begin{equation}
    \frac{\partial I^{\mathrm{sca}}(\theta)}{\partial \hat{b}_{n\ell}}=\frac{1}{Z_o}\gamma_{n\ell}\left(\sum_{m=1}^N\sum_{p=-\ell_{\mathrm{max}}}^{\ell_{\mathrm{max}}} \hat{b}_{mp}\gamma_{mp} \right)^*
\end{equation}

\begin{figure}[ht!]
	\centering
	\includegraphics[width=\linewidth]{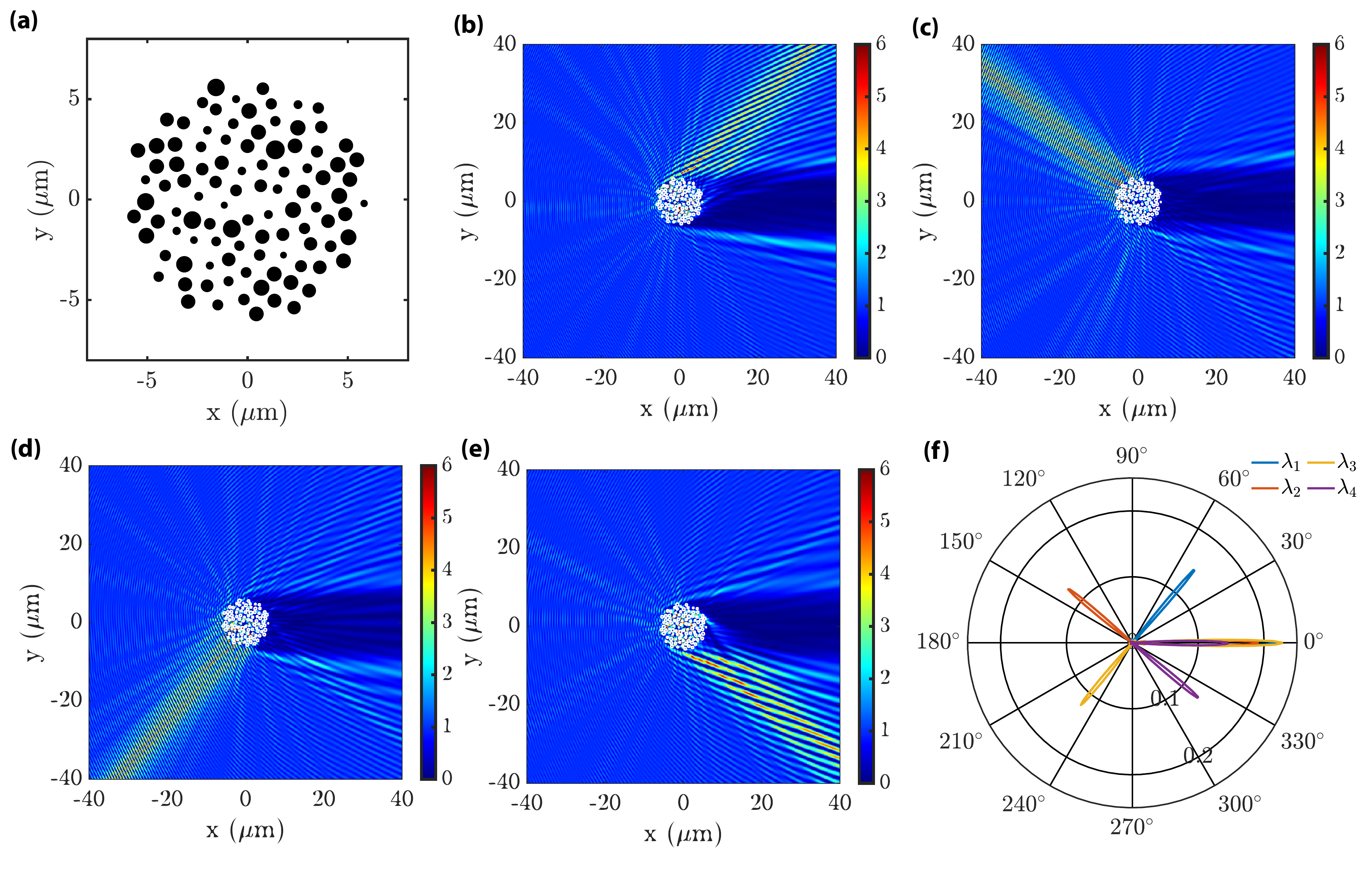}
	\caption{Example of a photonic patch optimized starting from the GA Vogel spiral geometry that steers four wavelengths into four desired directions. (a) Optimized photonic patch geometry. Total electric field intensity distributions under TM plane wave excitation at wavelengths (b) $\lambda_1=1.0\,\mu\mathrm{m}$, (c) $\lambda_2=1.1\,\mu\mathrm{m}$, (d) $\lambda_3=1.2\,\mu\mathrm{m}$, (e) $\lambda_4=1.3\,\mu\mathrm{m}$, respectively. (f) Polar plot of the scattered far-field intensities.}
	\label{fig:scattering_VS_4angles}
\end{figure}

Now we have computed all the analytical derivatives needed to perform the adjoint optimization of directional photonic patches within the framework of the 2D-GMT. These are compact photonic systems with optimized far-field scattered intensity at angle $\theta_o$ and wavelength $\lambda_o$. As a concrete demonstration of the developed method, we optimize for both the positions and the radii of the individual nanocylinders in a photonic patch in order to achieve simultaneous steering of radiation at wavelengths $\lambda_1$ and $\lambda_2$ at angles $\theta_1$ and $\theta_2$, respectively. Therefore, we introduce the objective function:
\begin{equation}\label{eq: radiation shaping obj function}
    \mathrm{g}=\frac{1}{I^{\mathrm{sca}}(\theta_1,\lambda_1)}+\frac{1}{I^{\mathrm{sca}}(\theta_2,\lambda_2)}
\end{equation}
We use the gradient descent method to update the design parameters in each iteration. Specifically, at the $k\mathrm{th}$ iteration, we have:
\begin{equation}
    \mathbf{p^\mathrm{k}} \leftarrow \mathbf{p^\mathrm{k-1}} - \alpha \frac{\partial \mathrm{g^{k-1}}}{\partial \mathbf{p^\mathrm{k-1}}}
\end{equation}
where $\alpha$ is the learning rate. The objective function value will decrease in each iteration and the far-field intensities $I^{\mathrm{sca}}(\theta_1,\lambda_1)$ and $I^{\mathrm{sca}}(\theta_2,\lambda_2)$ will increase by optimizing the design parameters. 
In our optimization, we chose $\lambda_1=1.0\,\mu\mathrm{m}$, $\lambda_2=1.1\,\mu\mathrm{m}$, $\theta_1=50^{\circ}$, and $\theta_2=70^{\circ}$ for the parameters of the objective function in Eq. \ref{eq: radiation shaping obj function}. The excitation was set to be TM polarized plane wave. We used a learning rate equal to $0.2$ for updating the cylinder radii and $0.02$ for updating their center positions. 
During our 2D-GMT calculations, the maximum angular order was set to $\ell_{\mathrm{max}}=3$, which is large enough to produce accurate results. 

We start from an initial array of $99$ cylinders arranged in the Vogel spiral structure, which is defined in polar coordinates as follows:
\begin{equation}\label{Vogel spiral}
    \begin{aligned} 
        \begin{cases}
            r_{n} &=a_{0} \sqrt{n} \\ \theta_{n} &=n \alpha 
        \end{cases}
    \end{aligned}
\end{equation}
where $n=0,1,2,...$ is an integer, $a_0$ is a positive constant called scaling factor, and $\alpha$ is an irrational number, known as the divergence angle \cite{adam2011mathematical}. Specifically for GA Vogel spirals $\alpha=360^\circ/\phi^2$, where $\phi=(1+\sqrt{5})/2\approx1.618$ is known as the golden number. The divergence angle determines the constant aperture between successive point particles in the array. Since it is an irrational number, Vogel spiral arrays lack both translational and rotational symmetry. Vogel spiral structures have been largely investigated in plasmonics and nanophotonics due to their unique light scattering and localization properties that enable compact photonic devices with broadband enhanced light-matter interactions \cite{trevino2012geometrical, lawrence2012control, pollard2009low, trevino2011circularly, liew2011localized, trevino2012plasmonic, razi2019optimization, fab2019localization, dalnegro2022waves}. 

\begin{figure}[ht!]
	\centering
	\includegraphics[width=\linewidth]{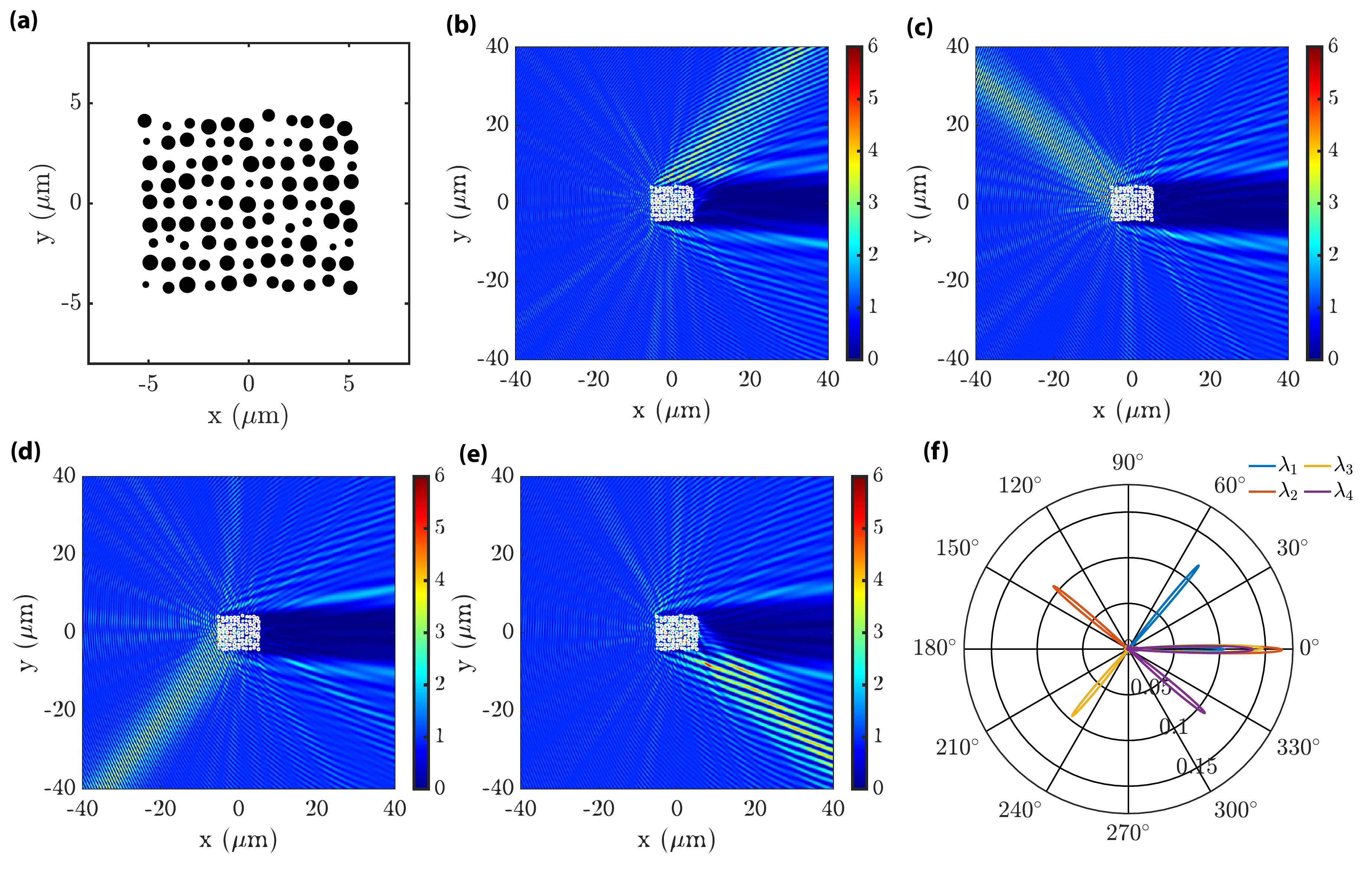}
	\caption{Example of a photonic patch optimized starting from a square array geometry that steers four wavelengths into four directions. (a) Optimized photonic patch geometry. Total electric field intensity distributions under TM plane wave excitation at wavelengths (b) $\lambda_1=1.0\,\mu\mathrm{m}$, (c) $\lambda_2=1.1\,\mu\mathrm{m}$, (d) $\lambda_3=1.2\,\mu\mathrm{m}$, (e) $\lambda_4=1.3\,\mu\mathrm{m}$, respectively. (f) Polar plot of the scattered far-field intensities. }
	\label{fig:scattering_periodic_4angles}
\end{figure}

In our simulations we considered  an initial GA Vogel spiral array with an averaged center-to-center particle separation  $\sim1~{\mu}\mathrm{m}$. We also set the initial cylinder radii $r=300\,\mathrm{nm}$, and the permittivity for the nanocylinder material $\epsilon=2.25$. We also took into consideration the practical limitations of our current fabrication technology and set the minimum radius for all cylinders to be no less than $50\,\mathrm{nm}$ during the optimization. A small value of the permittivity allows us to take advantage of long-range coupling effects across the entire array of nanocylinders, making its geometrical optimization more effective in this limit (see Fig. \ref{fig:eff_eps} for results obtained with different permittivity values). We display the initial array geometry in Fig. \ref{fig:scattering_VS}(a). We optimize over $200$ iterations and the value of the objective function during the optimization process is shown in Fig. \ref{fig:scattering_VS}(c). Convergence is obtained around $\mathrm{g}\sim 1$. Furthermore, the optimized array geometry is shown in Fig. \ref{fig:scattering_VS}(b). We observe that in the optimal array the cylinder radii are not all equal and the positions of each cylinder are also shifted with respect to the initial GA geometry. In Figs. \ref{fig:scattering_VS}(d) and \ref{fig:scattering_VS}(e) we display the spatial distributions of the total electric field under TM plane wave excitation at wavelengths $\lambda_1$ and $\lambda_2$, respectively. We also show in a polar plot the obtained scattered far-field intensities in Fig. \ref{fig:scattering_VS}(f) when the optimal structure is illuminated at wavelengths $\lambda_1$ and $\lambda_2$. The far-field pattern clearly demonstrates that the incident wavelengths $\lambda_1$ and $\lambda_2$ are steered at the desired angular directions $50^{\circ}$ and $70^{\circ}$, respectively. The differential scattering efficiencies of $\lambda_1, \lambda_2$, which are defined by the ratio of scattered light power along $\theta_1, \theta_2$ and their corresponding input power \cite{forouzmand2018tunable}, are estimated to be $19\%$ and $17\%$. Note that with a unit plane wave as the incident source, the differential scattering efficiency along a given angle is equal to the differential scattering cross-section that is dependent on directional angle \cite{gagnon_JO_2015}. The expression for the differential scattering cross-section is given in Eq. \ref{eq: differsca}. We remark that the  efficiency values that we have obtained in the optimized photonic patches are comparable to what has been achieved in beam steering applications using metasurface technologies \cite{berini2022optical, forouzmand2018tunable, shirmanesh2020electro, calalesina2021tunable, park2021allsolidstate}. 

We compared the optimization results of the GA Vogel spiral geometry with an 11$\times$9 periodic nanocylinder array characterized by the same averaged interparticle distance  $1~\mu\mathrm{m}$. The objective function parameters, learning rates and maximum multipole order $\ell$ are chosen to be the same as in the GA Vogel spiral case. We show the initial and optimized array geometries in Fig. \ref{fig:scattering_periodic}(a) and (b). Fig. \ref{fig:scattering_periodic}(c) shows the convergence achieved around $\mathrm{g}\sim 1$, similar to the case with GA Vogel spiral. Figs. \ref{fig:scattering_periodic}(d, e) show the spatial distribution of the total electric field at $\lambda_{1}$ and $\lambda_{2}$, respectively, and Fig. \ref{fig:scattering_periodic}(f) shows the polar plot representation of the computed far-field intensity patterns for both the wavelengths. The obtained differential scattering efficiencies at $\lambda_1, \lambda_2$ are $15\%$ and $13\%$, respectively. 

\begin{figure}[t!]
	\centering
	\includegraphics[width=\linewidth]{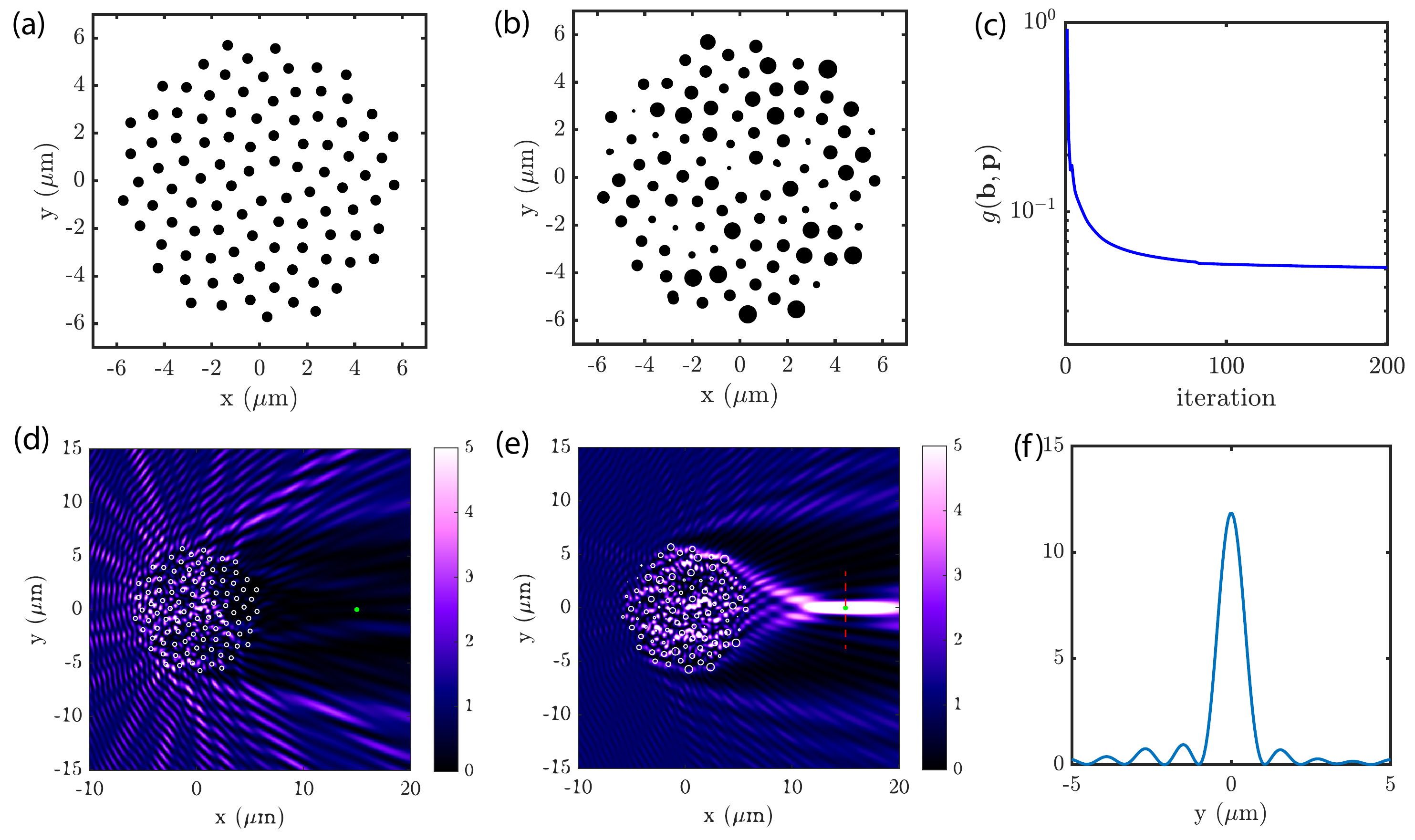}
	\caption{Example of a focusing photonic patch optimized starting from the GA Vogel spiral geometry. (a) Initial photonic patch geometry. (b) Optimized focusing patch geometry. (c) Objective function value with respect to the number of iterations. Total electric field intensity distributions for (d) initial and (e) optimized arrays under a TM plane wave excitation at $\lambda=1~\mu\mathrm{m}$. The green dot in (d) and (e) indicate the targeted focal position $(15\,\mu\mathrm{m},0)$. (f) The transverse profile of the focal spot along the red dashed line shown in panel (e).}
	\label{fig:focusing_VS}
\end{figure}

To further demonstrate the potential of our adjoint optimization method in the context of multi-wavelength radiation shaping, we optimize photonic patches that simultaneously steer incident waves at four different wavelengths into four desired far-field angles. The objective function that we used in this case is shown below: 
\begin{equation}\label{eq: radiation shaping obj function_4angles}
    \mathrm{g}=\sum_{i=1}^{4}\frac{1}{I^{\mathrm{sca}}(\theta_i,\lambda_i)}
\end{equation}
In this example we selected $\lambda_1=1.0\,\mu\mathrm{m}$, $\lambda_2=1.1\,\mu\mathrm{m}$, $\lambda_3=1.2\,\mu\mathrm{m}$, $\lambda_4=1.3\,\mu\mathrm{m}$, $\theta_1=50^{\circ}$, $\theta_2=140^{\circ}$, $\theta_3=230^{\circ}$, and $\theta_4=320^{\circ}$. The learning rate for updating radii and positions as well as $\ell_{\mathrm{max}}$ are kept the same as in the case of the previous optimizations. 
Similarly, we compare the results of an optimized GA Vogel spiral photonic patch to the ones of an optimized periodic array. The optimal GA Vogel spiral geometry is shown in Fig. \ref{fig:scattering_VS_4angles}(a). Moreover, Figs. \ref{fig:scattering_VS_4angles}(b) through \ref{fig:scattering_VS_4angles}(e) show respectively the total intensity distributions on the arrays at the four targeted wavelengths. The polar plot radiation diagram is displayed in Fig. \ref{fig:scattering_VS_4angles}(f) that demonstrates the ability of the optimized patch to steer incident radiation along the desired direction angles at each wavelength. The steering efficiencies at the four wavelengths are found to be $14\%, 13\%, 12\%, 13\%$. The corresponding results obtained by optimizing the periodic array are illustrated in Fig. \ref{fig:scattering_periodic_4angles}. In this case, the steering efficiencies at the four wavelengths were found to be $12\%, 11\%, 10\%, 11\%$. These results indicate that optimized photonic patches for multi-wavelength beam steering produce similar results regardless of the initial array geometry. 
Therefore, we have shown that our proposed approach can be used for the robust inverse design of photonics patches with small footprints that steer multiple wavelengths to desired directions. 

\begin{figure}[t!]
	\centering
	\includegraphics[width=\linewidth]{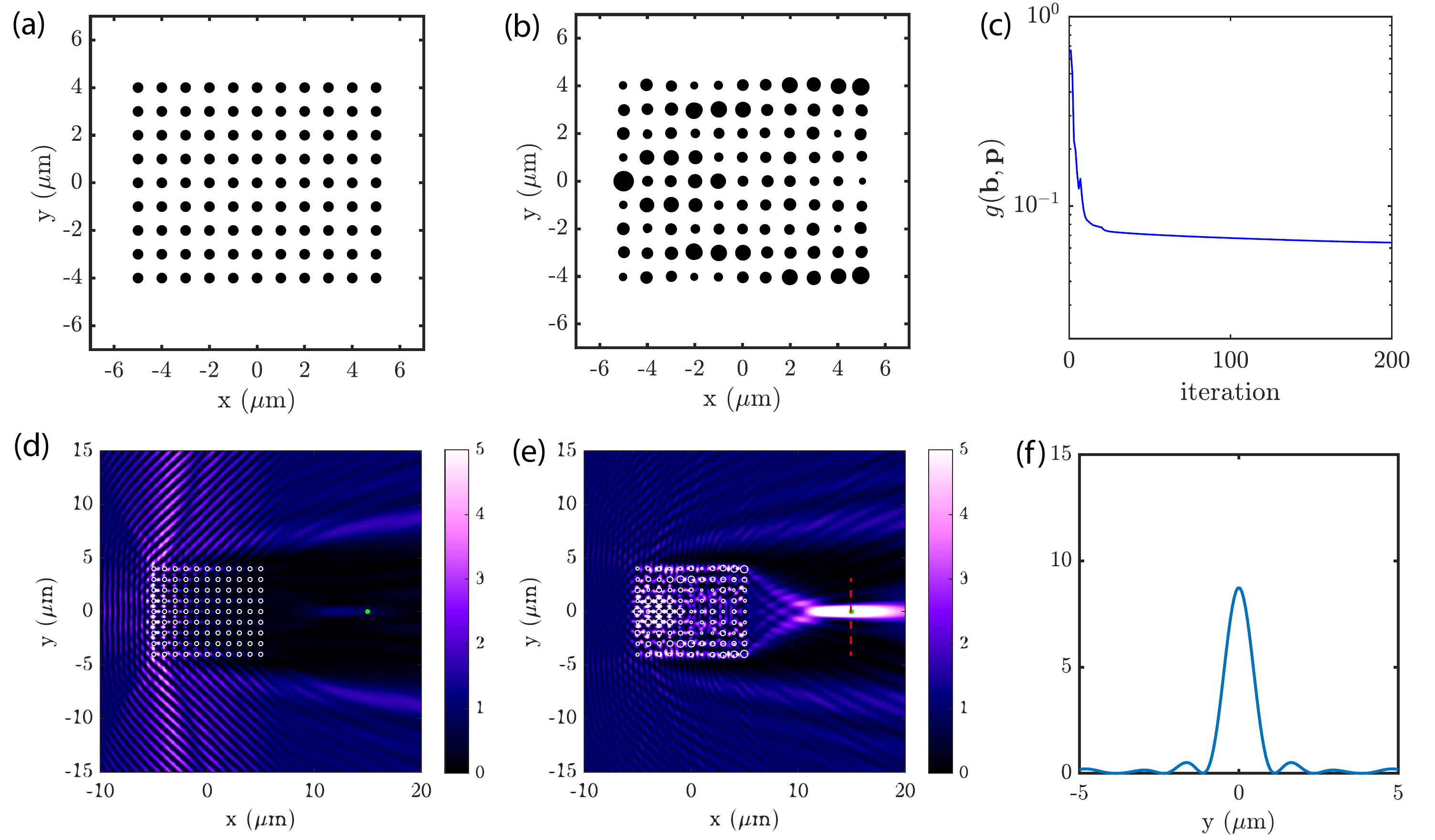}
	\caption{Example of a focusing photonic patch optimized starting from a periodic square array. (a) Initial photonic patch geometry. (b) Optimized focusing patch geometry. (c) Objective function value with respect to the number of iterations. Total electric field intensity distributions for (d) initial and (e) optimized arrays under a TM plane wave excitation at $\lambda=1~\mu\mathrm{m}$. The green dot in (d) and (e) indicates the targeted focal position $(15\,\mu\mathrm{m},0)$. (f) The transverse profile of the focal spot along the red dashed line shown in (e). }
	\label{fig:focusing_periodic}
\end{figure}

\subsection{Inverse design of photonic patches for radiation focusing}
\label{subsection: focusing photonic patch}
In this section we apply our inverse design methodology to optimize the focusing of incident radiation in the Fresnel zone using photonics patches. Specifically, we want to maximize the field intensity at a specific point $(x_f,y_f)$ under TM plane wave excitation for a generic wavelength $\lambda_o$. Our objective function is therefore:
\begin{equation}
    \label{eq:gIsca2}
    \mathrm{g}=\frac{1}{I^{\mathrm{sca}}_z(x_f,y_f;\lambda_o)}=\frac{1}{|\varphi^{\mathrm{sca}}_z(x_f,y_f;\lambda_o)|^2}
\end{equation}
where $\varphi^{\mathrm{sca}}_z \equiv \varphi_{z}^{E,\mathrm{sca}}$ was defined in Eq. \ref{eq:varphiEsca}. 
To enable the optimization of the focusing properties we need to compute the partial derivative of $\mathrm{g}$ with respect to the design parameters $p_i$:
\begin{equation}
    \frac{\partial \mathrm{g}}{\partial p_{i}}=-\frac{2}{(I^{\mathrm{sca}}_z(x_f,y_f;\lambda_o))^2} \Re\left\{\frac{\partial \varphi_{z}^{\mathrm{sca}}}{\partial p_{i}} \left(\varphi_{z}^{\mathrm{sca}}\right)^{*}\right\}
\end{equation}
where we have:
\begin{eqnarray}
    \frac{\partial \varphi_{z}^{\mathrm{sca}}}{\partial p_{i}}=\sum_{n \ell} \hat{b}_{n \ell} \tau_{n \ell}&& \bigg[k_{o} \frac{J_{\ell}^{\prime} \left(k_{o} r_{n}\right)}{J_{\ell}\left(k_{o} r_{n}\right)}  \frac{\partial r_{n}}{\partial p_{i}}  \nonumber\\
    && +  k_{o} \frac{H_{\ell}^{\prime}\left(k_{o} \rho_{n}\right)}{H_{\ell}\left(k_{o} \rho_{n}\right)} \frac{\partial \rho_{n}}{\partial p_{i}}+j \ell \frac{\partial \theta_{n}}{\partial p_{i}}\bigg]
\end{eqnarray}
and we defined $\tau_{n \ell}=p_{z} J_{\ell}\left(k_{o} r_{n}\right) H_{\ell}\left(k_{o} \rho_{n}\right) e^{j \ell \theta_{n}}$. Furthermore, the derivative of $\mathrm{g}_\mathbf{\hat{b}}$ can be computed as follows:
\begin{equation}
    \frac{\partial \mathrm{g}}{\partial \hat{b}_{n\ell}}=2 \Re\left\{\frac{\partial \varphi_{z}^{\mathrm{sca}}}{\partial \hat{b}_{n\ell}} \left(\varphi_{z}^{\mathrm{sca}}\right)^{*}\right\}
\end{equation}
where:
\begin{equation}\label{eq:phi_sca_grad_bnl}
    \frac{\partial\varphi_z^{\mathrm{sca}}}{\partial\hat{b}_{n\ell}}=\tau_{n\ell}\left(\varphi_{z}^{\mathrm{sca}}\right)^{*}
\end{equation}

\begin{figure}[t!]
	\centering
	\includegraphics[width=\linewidth]{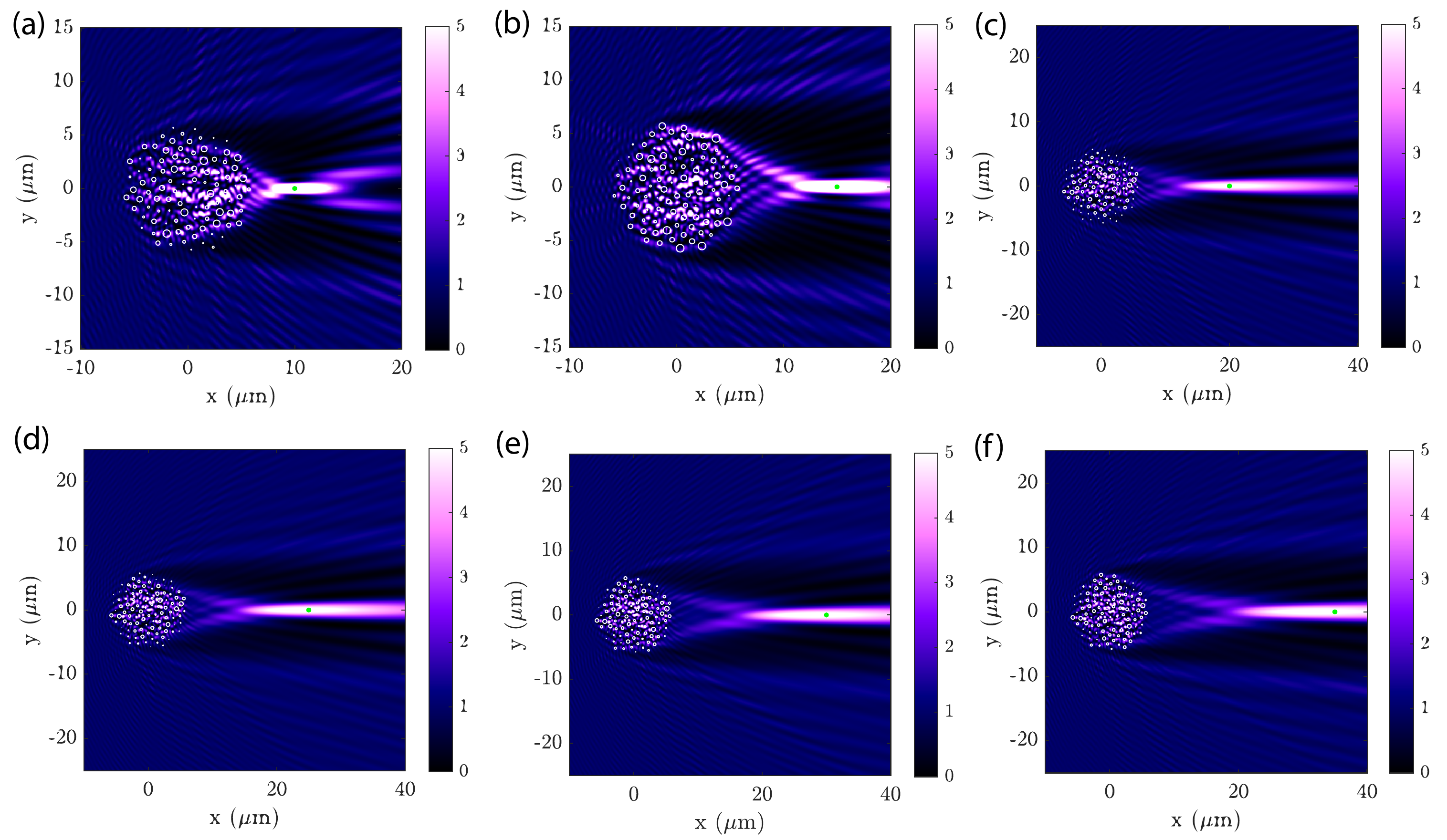}
	\caption{Six different optimized photonic patches obtained starting from the GA spiral geometry. These patches  focus incident light at the desired focal positions $x_f$ given by (a) $10~\mu\mathrm{m}$, (b) $15~\mu\mathrm{m}$, (c) $20~\mu\mathrm{m}$, (d) $25~\mu\mathrm{m}$, (e) $30~\mu\mathrm{m}$, (f) $35~\mu\mathrm{m}$. The $y_f$ coordinates for all cases are 0. The considered wavelength for the incident plane wave is $1\,\mu\mathrm{m}$.}
	\label{fig:focusing_VS_sixfls}
\end{figure}

In our focusing simulation we chose $\lambda=1\,\mu\mathrm{m}$, and $(x_f, y_f)=(15\,\mu\mathrm{m}, 0)$. We start from an initial array with 99 cylinders arranged in GA Vogel spiral, the same condition as in subsection \ref{section: adjoint 2D-GMT}\ref{subsection: AO radiation shaping}. We set the initial cylinder radii $r=200\,\mathrm{nm}$ and fix the permittivity of the nanocylinders material to be $\epsilon=2.25$. The maximum angular order is chosen as $\ell_{\mathrm{max}}=4$ in order to improve the accuracy in the near-field zone. We display the initial cylinder array geometry in Fig. \ref{fig:focusing_VS}(a). We use the same learning rates to update radii and positions as in the subsection \ref{section: adjoint 2D-GMT}\ref{subsection: AO radiation shaping}. We optimize the radii and centers of the cylinders in the array using $200$ iterations. The  optimized array geometry of the patch is shown in Fig. \ref{fig:focusing_VS}(b), where we clearly observe that the positions and radii of cylinders have been modified from the ones in the initial structure. The objective function with respect to the number of iterations is shown in Fig. \ref{fig:focusing_VS}(c). Furthermore, we show the total field intensities for both the initial and the optimized arrays under plane wave excitation at wavelength $\lambda$ in Fig. \ref{fig:focusing_VS}(d) and \ref{fig:focusing_VS}(e), respectively. The profile of the focal spot along the $x_f=15\,\mu\mathrm{m}$ line is shown in Fig. \ref{fig:focusing_VS}(f). The transverse full-width-at-half-maximum (FWHM) of the focusing spot is $0.98\,\mu\mathrm{m}$. The focusing efficiency, which is defined as the ratio between the power contained in the main lobe of the focal spot and that of the incident power deposited  on the area of the device, is calculated to be $77\%$. 

\begin{figure}[t!]
	\centering
	\includegraphics[width=\linewidth]{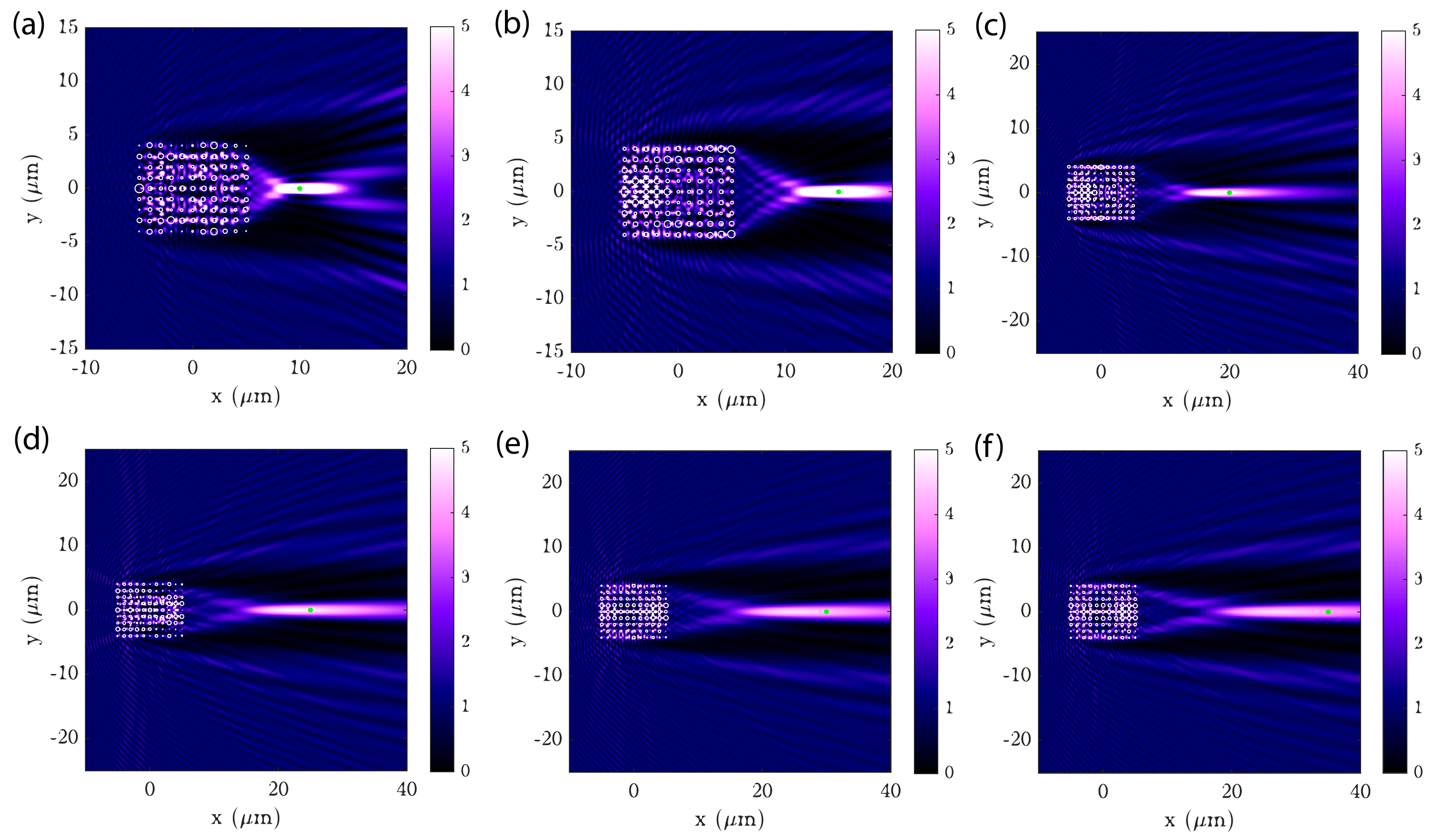}
	\caption{Six different optimized photonic patches obtained starting from the square array geometry. These patches  focus incident light at the desired focal positions $x_f$ given by (a) $10~\mu\mathrm{m}$, (b) $15~\mu\mathrm{m}$, (c) $20~\mu\mathrm{m}$, (d) $25~\mu\mathrm{m}$, (e) $30~\mu\mathrm{m}$, (f) $35~\mu\mathrm{m}$. The $y_f$ coordinates for all cases are 0. The considered wavelength for the incident plane wave is $1\,\mu\mathrm{m}$.}
	\label{fig:focusing_periodic_sixfls}
\end{figure}

As a comparison, we also optimized a 99-cylinder array starting from a periodic square structure. Keeping all the parameters the same as in the case of the GA Vogel spiral simulation, we display the initial and optimized array geometries in Fig. \ref{fig:focusing_periodic}(a) and \ref{fig:focusing_periodic}(b). Fig. \ref{fig:focusing_periodic}(c) shows the values of the objective function with respect to the number of iterations. Similarly to the case of the GA Vogel spiral, Fig. \ref{fig:focusing_periodic}(d) and \ref{fig:focusing_periodic}(e) display the spatial distributions of the total field intensity for the initial and the optimized array geometries respectively. In Fig. \ref{fig:focusing_periodic}(f) we illustrate the transverse profile of the focal spot along the $x_f=15\,\mu\mathrm{m}$ dashed line. The transverse FWHM of the focusing spot is also  $0.98\,\mu\mathrm{m}$. We can clearly observe that the field intensity at the desired location (indicated by the red dot) is strongly enhanced. The intensity profile is similar to that of a focusing lens. However, we emphasize that here we are achieving such a focusing behavior using an array of cylinders with a total dimension of $\sim10\,\mu\mathrm{m}$ and a focal length $x_f=15\,\mu\mathrm{m}$, which are challenging to obtain using traditional diffractive elements. Moroever, we found that the focusing efficiency of the optimized periodic patch is  $60\%$. It is noteworthy to observe that if one considers a diffraction-limited lens with the same diameter and dimension of the optimized photonic patch as well as the same focal length as $15\,\mu\mathrm{m}$, then the FWHM of at the focal spot according to Rayleigh criterion will be $1.58\,\mu\mathrm{m}$ \cite{goodman1996introduction}. This behavior reflects the structural complexity of the optimized aperiodic geometries of the patches which, analogously to what recently reported in random media \cite{vellekoop2007focusing, vellekoop2010exploiting}, produce a focal spot with significantly smaller FWHM compared to the traditional Rayleigh diffraction limit. 

\begin{figure}[t!]
	\centering
	\includegraphics[width=\linewidth]{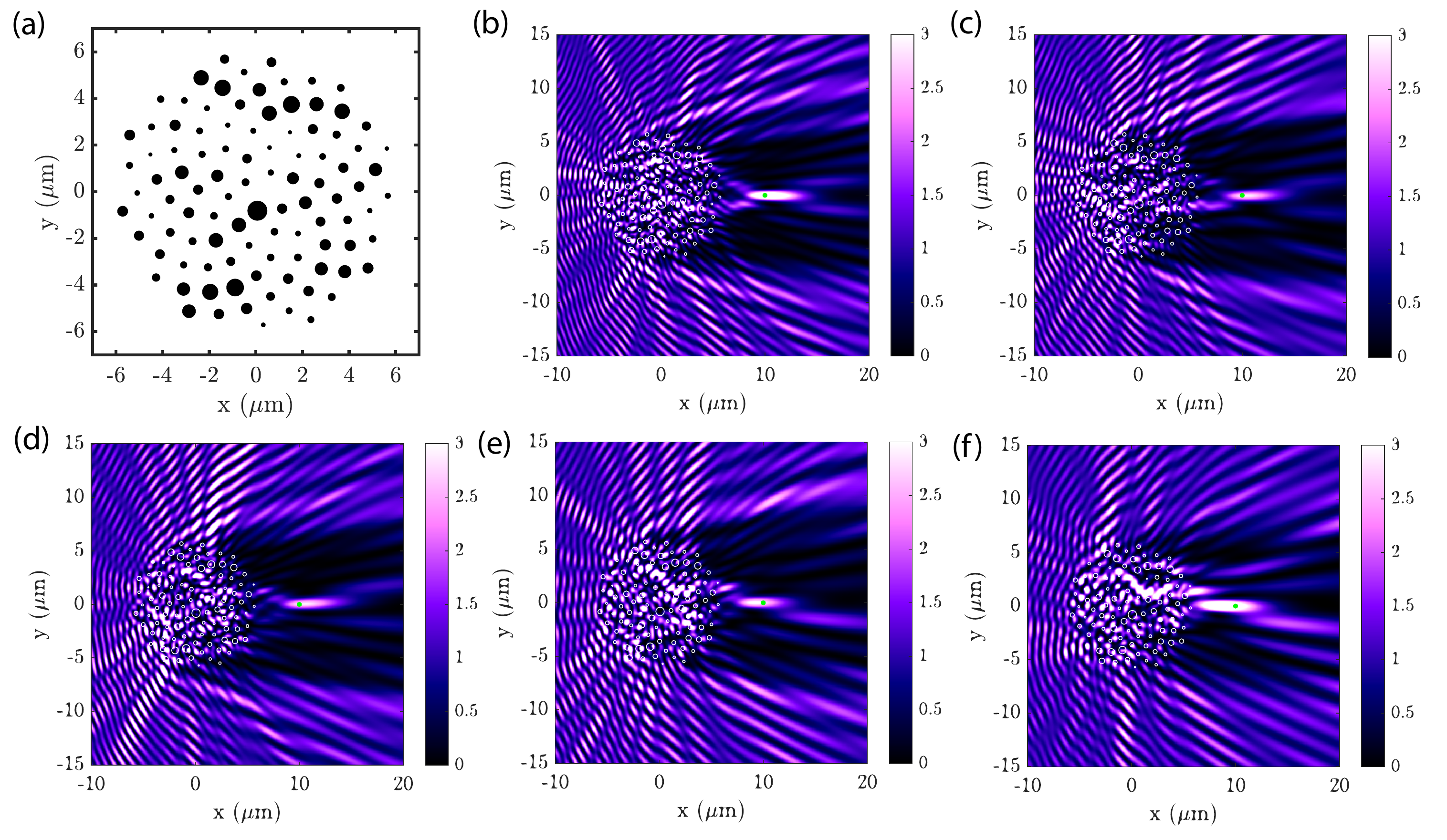}
	\caption{(a) Optimized GA Vogel spiral geometry for achromatic focusing at $x_f=10\,\mu\mathrm{m}$. Also shown are the field intensity distributions with incident wavelengths (b) $\lambda_1=1.0~\mu\mathrm{m}$, (c) $\lambda_2=1.1~\mu\mathrm{m}$, (d) $\lambda_3=1.2~\mu\mathrm{m}$, (e) $\lambda_4=1.3~\mu\mathrm{m}$, (f) $\lambda_5=1.4~\mu\mathrm{m}$. }
	\label{fig:focusing_VS_ach}
\end{figure}

\begin{figure}[t!]
    \centering
    \includegraphics[width=\linewidth]{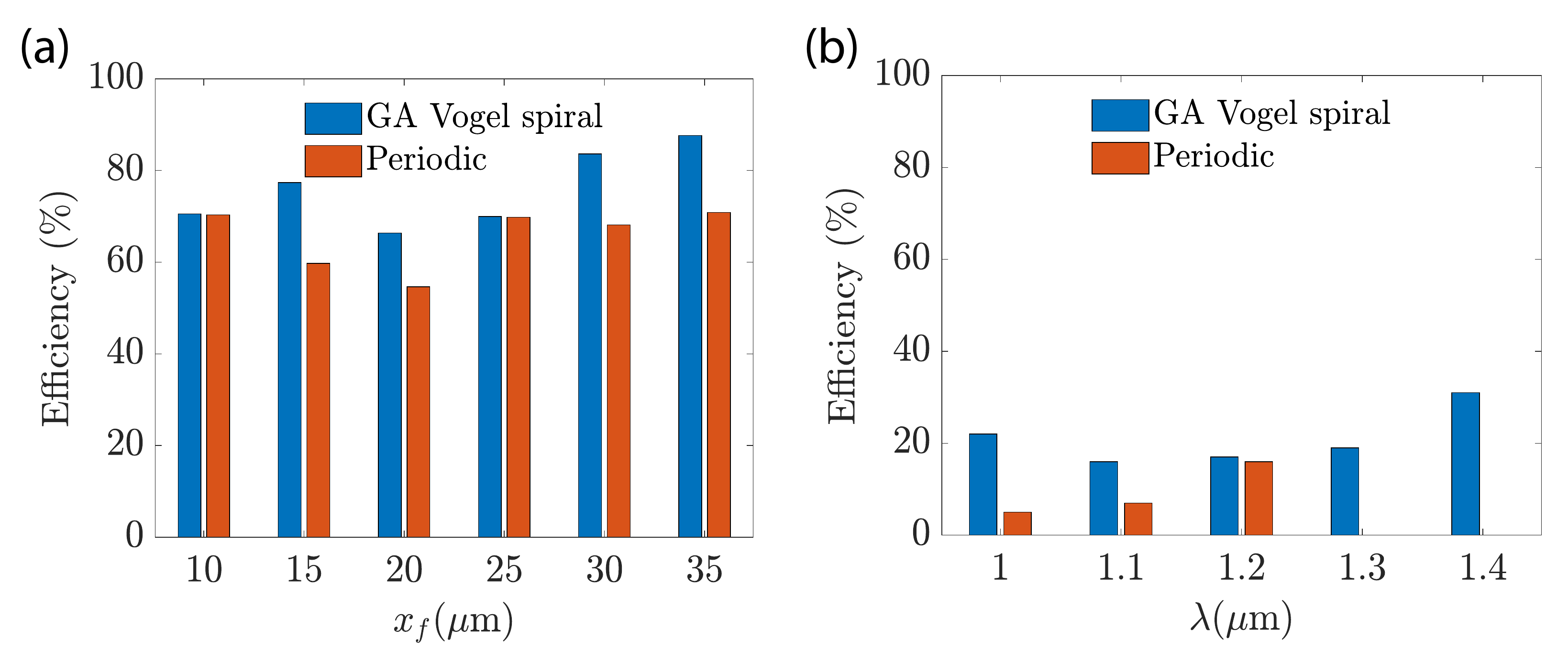}
    \caption{(a) Single wavelength ($\lambda=1\,\mu\mathrm{m}$) focusing efficiencies for optimized GA Vogel spirals (blue) and periodic array (red) photonic patches with different focal positions $x_f$.  (b) Broadband focusing efficiencies of optimized GA Vogel spiral (blue) and periodic array (red) achromatic patches at different incident wavelengths $\lambda$. The focusing positions are all $(x_f,y_f)=(10\,\mu\mathrm{m},0)$. }
    \label{fig:focusing_efficiencies}
\end{figure}

We further optimized photonic patches that are able to focus incident light at different focal lengths. Fig. \ref{fig:focusing_VS_sixfls} and \ref{fig:focusing_periodic_sixfls} show the total field intensity patterns of six different devices with focal positions at $x_f=10,15,20,25,30,35\,\mu\mathrm{m}$, for the initial GA Vogel spiral geometry and periodic geometry, respectively. The incident wavelength for all devices is $\lambda=1\,\mu\mathrm{m}$. The focusing efficiencies obtained for the GA Vogel spiral and periodic geometry, for different focal positions $x_f$ are listed in Fig. \ref{fig:focusing_efficiencies}(a). Our results indicate that when considering only one focusing wavelength, the focusing efficiencies of the optimized GA Vogel spiral patches and periodic patches are quite comparable. However, it is also very relevant to consider the case of broadband incident radiation, which we address below.

\begin{figure}[t!]
    \centering
    \includegraphics[width=\linewidth]{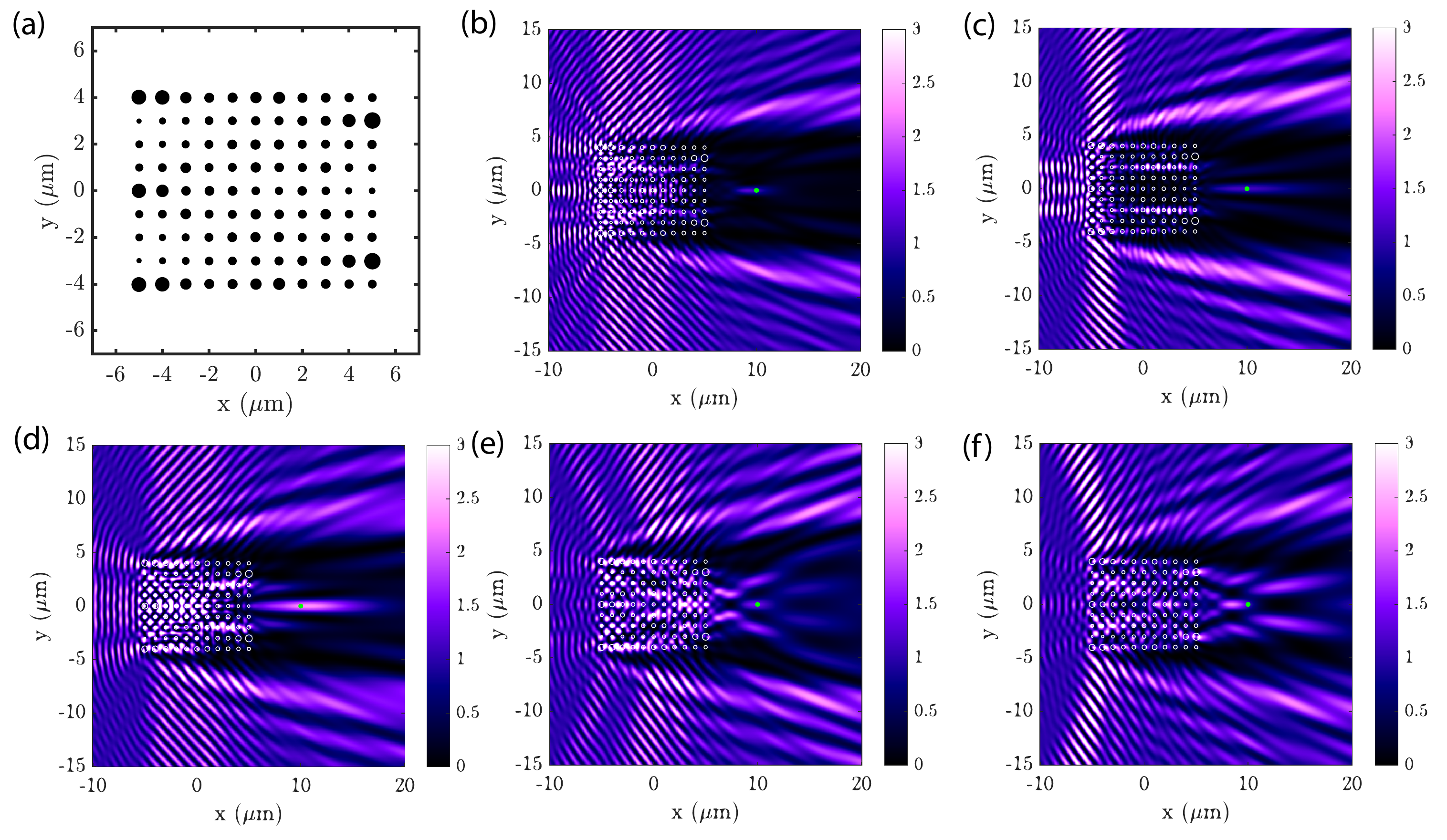}
    \caption{(a) Optimized periodic array photonic patch for achromatic focusing at $x_f=10\,\mu\mathrm{m}$. Also shown are the field intensity distributions with incident wavelengths (b) $\lambda_1=1.0~\mu\mathrm{m}$, (c) $\lambda_2=1.1~\mu\mathrm{m}$, (d) $\lambda_3=1.2~\mu\mathrm{m}$, (e) $\lambda_4=1.3~\mu\mathrm{m}$, (f) $\lambda_5=1.4~\mu\mathrm{m}$. }
    \label{fig:focusing_pd_ach}
\end{figure}

The goal is to investigate the inverse design of optimized photonic patches for broadband focusing applications. This can be achieved by considering the multi-objective function defined as:
\begin{equation}
    \label{eq:gIscaach}
    \mathrm{g}=\sum_{i=1}^{5}\frac{1}{I^{\mathrm{sca}}_z(x_f,y_f;\lambda_i)}+\sum_{i\ne j}\left[I^{\mathrm{sca}}_z(x_f,y_f;\lambda_i)-I^{\mathrm{sca}}_z(x_f,y_f;\lambda_j)\right]^2
\end{equation}
Note that the summation over all wavelengths ensures that the focal spot intensities at multiple incident wavelengths are mutually maximized. Moreover, to prevent the situation where the focal intensity of only one wavelength is maximized, we introduced above a cross term that penalizes large focal intensity differences for any pair of distinct wavelengths. To illustrate the approach, we selected the five incident wavelengths  $\lambda_1=1.0\,\mu\mathrm{m}, \lambda_2=1.1\,\mu\mathrm{m}, \lambda_3=1.2\,\mu\mathrm{m}, \lambda_4=1.3\,\mu\mathrm{m}, \lambda_5=1.4\,\mu\mathrm{m}$ and the focal position was chosen to be $(x_f,y_f)=(10\,\mu\mathrm{m},0)$. We optimized the patches using a learning rate of 0.2 for updating cylinder radii and that of 0.02 for updating cylinder positions, and the total number of iterations used here was 1000. Fig. \ref{fig:focusing_VS_ach} illustrates the total field intensity distributions of the device optimized starting from a GA Vogel spiral geometry, while Fig. \ref{fig:focusing_pd_ach} illustrates those of the optimized periodic devices. The focusing efficiencies obtained for the GA Vogel spiral and periodic geometries are compared at the different incident wavelengths $\lambda$ in Fig. \ref{fig:focusing_efficiencies}(b). Note that the optimized periodic patch does not appreciably focus incident light at wavelengths $\lambda_4=1.3\,\mu\mathrm{m}$ and $\lambda_5=1.4\,\mu\mathrm{m}$, and therefore the focusing efficiencies could not be defined for those values. 
From our analysis we conclude that while the focusing performances at a single wavelength are comparable for the two considered geometries, the optimized GA aperiodic patches show significant efficiency advantages in broadband focusing compared to the optimized periodic arrays.

\begin{figure}[t!]
    \centering
    \includegraphics[width=\linewidth]{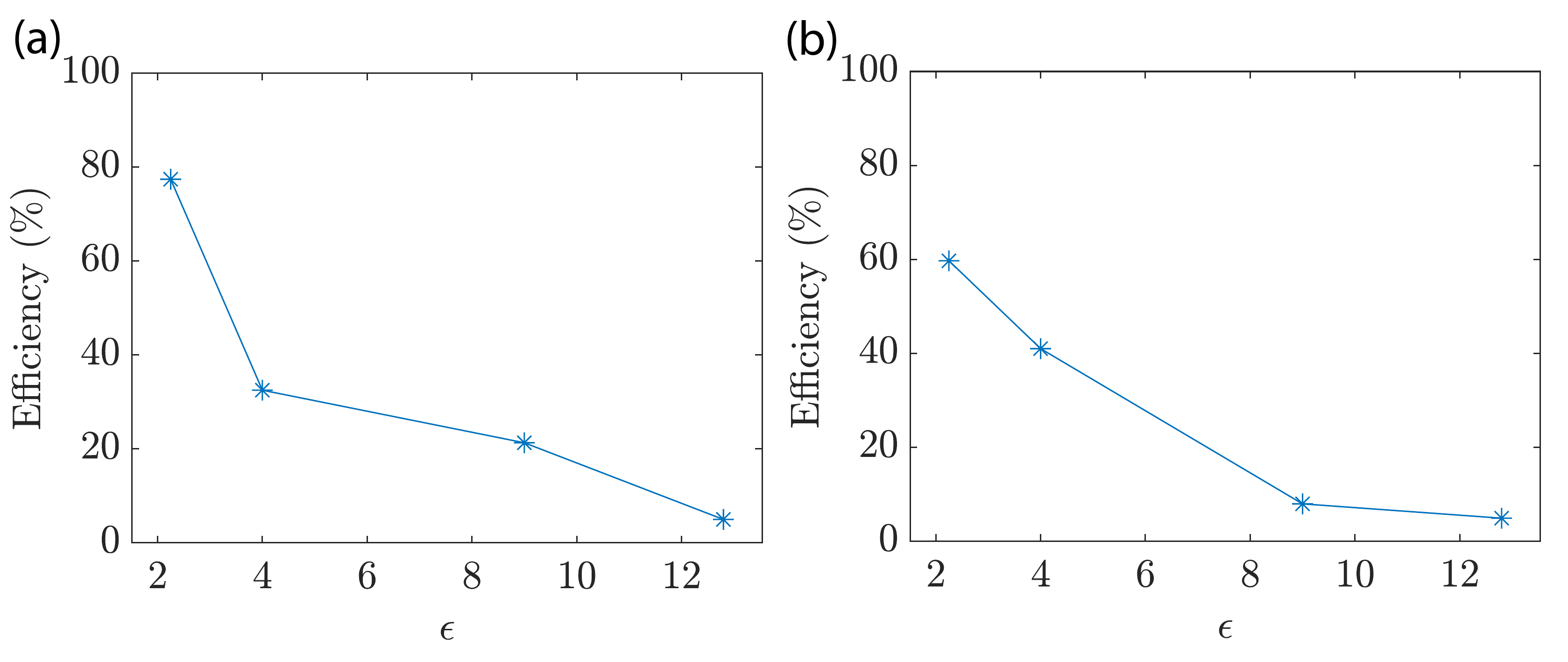}
    \caption{Focusing efficiencies as a function of the permittivity $\epsilon$ of the nanocylinders for photonic patches with (a) initial GA Vogel spiral geometry and (b) initial periodic array geometry. The focusing position is $x_f=15\,\mu\mathrm{m}, y_f=0$, and the wavelength is $\lambda=1\,\mu\mathrm{m}$. }
    \label{fig:eff_eps}
    \end{figure}

We finally characterized the focusing efficiency of photonic patches as a function of the permittivity $\epsilon$ of the dielectric cylinders. We directly compared arrays optimized starting from the GA Vogel spiral and the square array configurations, using the same structural parameters as in Figs. \ref{fig:focusing_VS}(a) and \ref{fig:focusing_periodic}(b). The focal distance was set to $x_f=15\,\mu\mathrm{m}, y_f=0$ and the wavelength is $\lambda=1\,\mu\mathrm{m}$. Our results are shown in Figs. \ref{fig:eff_eps}(a) and (b) that demonstrate how for both configurations the focusing efficiencies decrease when $\epsilon$ is increased. This behavior reflects the more localized nature of the resonances supported for larger $\epsilon$, reducing long-range electromagnetic coupling and the effectiveness of the geometrical optimization in this limit \cite{krauss1999photonic,wiesmann2009photonic,lawrence2012aperiodic}. 

\subsection{Scaling analysis of photonic patches}\label{subsection: efficiency Nparticles}
\begin{figure}[t!]
	\centering
	\includegraphics[width=\linewidth]{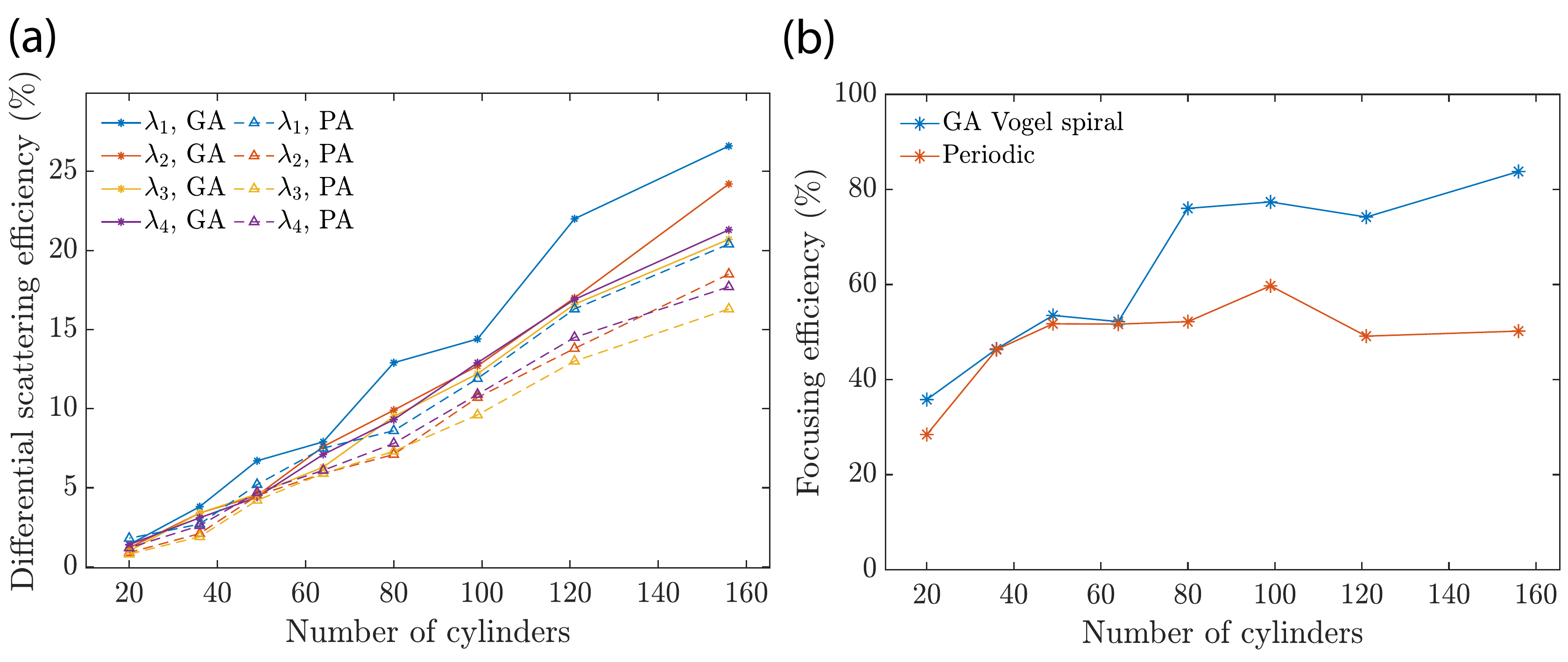}
	\caption{(a) Differential scattering efficiencies for each wavelength (color labeled in the legend) versus the number of cylinders in the photonics patches with initial GA Vogel spiral geometry (labelled as "GA") and initial periodic array geometry (labeled as "PA") geometries. (b) Focusing efficiencies of the optimized GA Vogel spirals (blue) and optimized periodic arrays (red). }
	\label{fig:Efficiency_Nparticles}
\end{figure}

A key question in the design of photonic patches is related to what is the smallest size of the array that still achieves a desired functionality. To answer that question, we systematically investigated how the performance of proposed photonic patches scales with the overall footprint of the scattering array. Specifically, we varied the number of nanocylinders $N$ designated in beam shaping and focusing patches and then simulated their corresponding far-field steering or focusing efficiencies. Fig. \ref{fig:Efficiency_Nparticles}(a) shows the results of the differential scattering efficiency of the optimized photonic patches at different wavelengths $\lambda_1=1.0\,\mu\mathrm{m}, \lambda_2=1.1\,\mu\mathrm{m}, \lambda_3=1.2\,\mu\mathrm{m}, \lambda_4=1.3\,\mu\mathrm{m}$ versus $N$. As one can see, the differential scattering efficiency at each wavelength changes almost linearly with respect to $N$ for both GA Vogel spiral (shown in solid lines) and periodic structures (shown in dashed lines) for all the wavelengths (labeled by the different colors). However, our results indicate that  the optimization of aperiodic GA Vogel spirals produces higher efficiencies compared to the one of  periodic arrays. Fig. \ref{fig:Efficiency_Nparticles}(b) shows the focusing efficiencies at $x_f=15\,\mu\mathrm{m}$ with incident wavelength $\lambda=1\,\mu\mathrm{m}$ versus $N$, for both optimized GA Vogel spiral and periodic structures. The results show that the focusing efficiencies saturate beyond a critical number of cylinders for both periodic and aperiodic structures. However, we also observe that the optimal focusing efficiencies that can be obtained by optimizing  the GA Vogel spiral structures are consistently larger than the ones possible with optimized periodic arrays.

\subsection{Inverse design of photonic patches for LDOS enhancement}\label{subsection: LDOS}
We address in this subsection the enhancement of the LDOS in optimized GA aperiodic photonic patches through adjoint optimization. Depending on the orientation of the excitation dipole, the Purcell enhancement depends on the $\Im\left\{\mathrm{G}_{zz}\right\}$ only, for TM polarization, or on the $\Im\left\{\mathrm{G}_{xx}+\mathrm{G}_{yy}\right\}$, for TE polarization. Therefore, we introduce the following objective function for maximizing the TE Purcell enhancement:
\begin{equation}\label{Eq:gTE}
    \mathrm{g}_\mathrm{TE} = \Im\left\{\mathrm{G}_{xx}+\mathrm{G}_{yy}\right\}
\end{equation}
and for the TM Purcell enhancement we use:
\begin{equation}\label{Eq:gTM}
    \mathrm{g}_\mathrm{TM} = \Im\left\{\mathrm{G}_{zz}\right\}
\end{equation}
where the expressions for $\mathrm{G}_{xx}$, $\mathrm{G}_{yy}$, and $\mathrm{G}_{zz}$ were given in Eqs. \ref{Eq:Gxx} through Eqs. \ref{Eq:Gzz}. The derivative $\mathrm{g}_{\mathbf{\hat{b}}}$ can be readily obtained from the following expressions:
\begin{eqnarray}
    \frac{\partial\left(\Im G_{x x}\right)}{\partial \hat{b}_{n \ell}}= && b_{n \ell} e^{j \ell \theta_{n s}}\bigg[H_{\ell}^{\prime}\left(k_{o} R_{n s}\right) \sin \left(\theta_{n s}\right) \nonumber\\
    &&\left. +\frac{j \ell}{k_{o} R_{n s}} H_{\ell}\left(k_{o} R_{n s}\right) \cos \left(\theta_{n s}\right)\right]\\
    \frac{\partial\left(\Im G_{y y}\right)}{\partial \hat{b}_{n \ell}}= && -b_{n \ell} e^{j \ell \theta_{n s}}\bigg[H_{\ell}^{\prime}\left(k_{o} R_{n s}\right) \cos \left(\theta_{n s}\right)\nonumber\\
    && \left. -\frac{j \ell}{k_{o} R_{n s}} H_{\ell}\left(k_{o} R_{n s}\right) \sin \left(\theta_{n s}\right)\right]\\
    \frac{\partial\left(\Im G_{z z}\right)}{\partial \hat{b}_{n \ell}}= && -j J_{\ell}\left(k_{o} r_{n}\right) H_{\ell}\left(k_{o} R_{n s}\right) e^{j \ell \theta_{n s}}
\end{eqnarray}

\begin{figure}[t!]
    \centering
    \includegraphics[width=\linewidth]{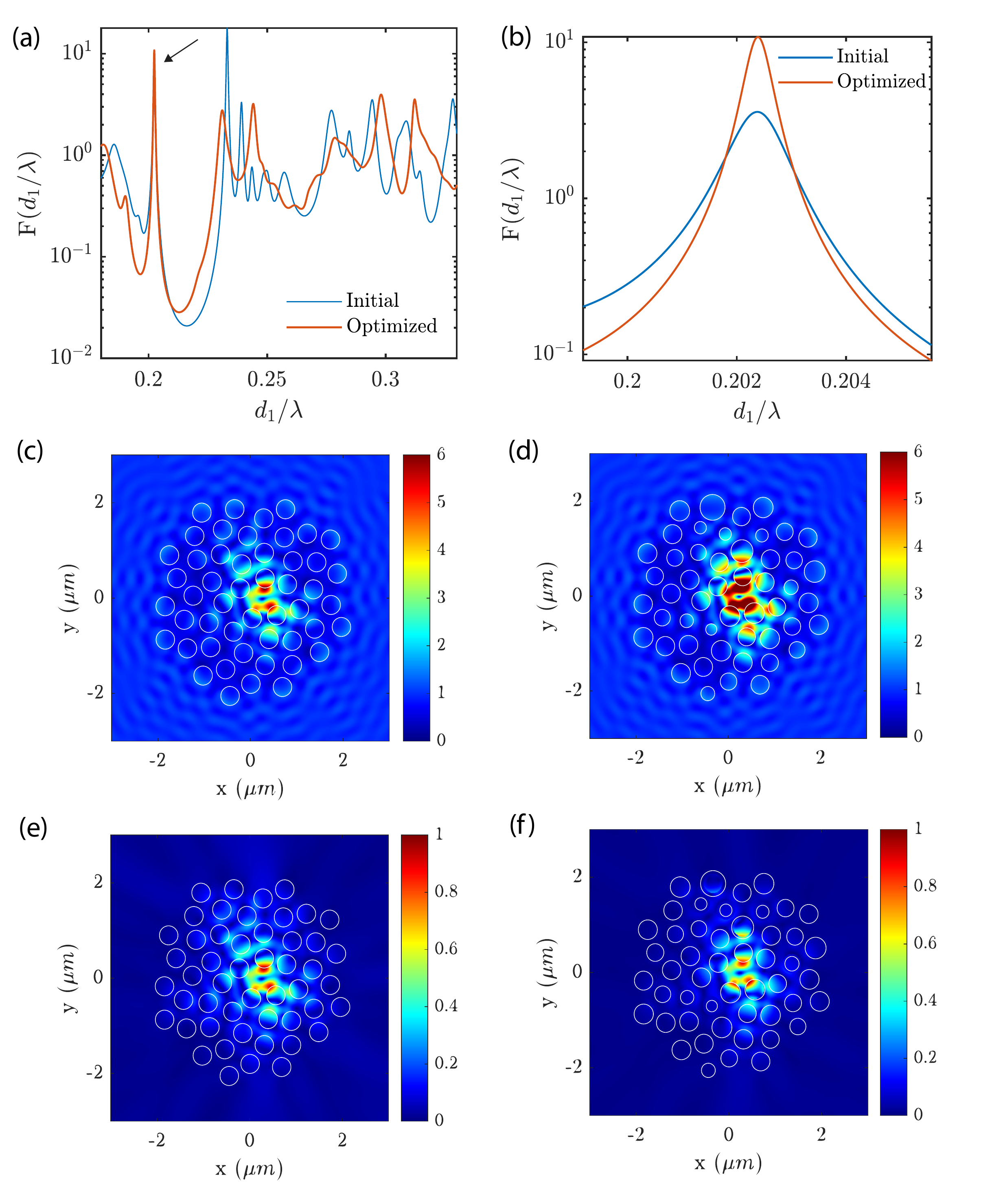}
    \caption{(a) Purcell factor spectrum for both the initial (blue) and optimized (red) GA Vogel spiral air-hole array with TE polarized dipole placed at $(0.4\,\mu\mathrm{m}, 0.02\,\mu\mathrm{m})$. The black arrow indicates the spectral parameter $d_1/\lambda=0.202363$ where we performed adjoint optimization. (b) Purcell factor spectrum for the initial (blue) and optimized (red) photonic patch near the optimized mode. The spatial map of Purcell factors of (c) initial and (d) optimized photonic patches at $d_1/\lambda=0.202363$. Also shown are the spatial distributions of TE-polarized optical modes for the (e) initial and (f) optimized photonic patches, respectively.}
    \label{fig:LDOS_TE}
\end{figure}

As a relevant example of LDOS enhancement in small-size photonic patches we consider first the optimization of the Purcell factor for arrays of dielectric cylinders arranged initially in the GA Vogel spiral geometry. In Fig.\,\ref{fig:LDOS_TE} we present an example related to the optimization of a TE mode of the GA Vogel structure. The initial array consists of a GA Vogel spiral with only 50 air holes of initial radii $r=200\,\mathrm{nm}$ and averaged center-to-center particle distance $d_1=0.50~{\mu}\mathrm{m}$. Note that here the cylinders consist of air-holes embedded in a dielectric medium with $\epsilon_o=12.8$, since in this configuration a TE-polarized bandgap is expected to open for relatively small-size arrays, hosting high-quality factor band-edge modes \cite{oliver2020cavity, oliver2021cavity,dal2021aperiodic}. In particular, the analyzed  structure supports a strong band-edge resonance excited by a dipole with in-plane orientation for  $d_1/\lambda=0.202363$. The excitation dipole is located  at position $(x_s,y_s)=(0.4\,\mu\mathrm{m}, 0.02\,\mu\mathrm{m})$ and we maximize the TE Purcell enhancement at this band-edge resonance by adjusting all the radii and positions of the cylinders in the array. In our computation we selected a learning rate for updating the radii equal to $0.1$ while the one used for updating the positions was set equal to $0.001$. We then optimized the array using $20000$ iterations. In Fig. \ref{fig:LDOS_TE}(a) we compare the Purcell enhancement spectrum for both the initial and the optimized geometry of the array. The black arrow in the panel indicates the spectral position of the targeted band-edge mode. Fig. \ref{fig:LDOS_TE}(b) clearly illustrates the significant enhancement achieved for the Purcell factor of the considered resonant mode. Moreover, the Purcell enhancements for the initial and the optimized arrays are found to be $\mathrm{F}_i\approx3.58$ and $\mathrm{F}_o\approx27.5$, resulting in an increase by a factor of $7.67$ due to the reduced mode volume of the optimized resonance. We further  characterized the optical resonant modes by solving the homogeneous T-matrix equation $\mathbf{Tb}=0$. The resonant modes are obtained by finding the complex eigenvalues  $k=\Re(k)+j\Im(k)$ that satisfy the relation $\det[\mathbf{T}(k)]=0$ \cite{gagnon_JO_2015,dal2021aperiodic}. Here, $\Re(k)$ is equal to the wavenumber of the mode, while $\Im(k)$ corresponds to its decay rate, which is inversely proportional to the spectral width of the mode. We evaluated the resonant modes by generating a 2D map of $\det[\mathbf{T}(k)]$ with a resolution of $\Delta[\Re(k)]=1.25\times10^{-4}\,{\mu\mathrm{m}}^{-1}$ and $\Delta[\log_{10}\Im(k)]=0.05$ \cite{dal2021aperiodic}. The corresponding quality factors are computed according to $\mathrm{Q}=\abs{\Re(k)/[2\Im(k)]}$ \cite{gagnon_JO_2015, dal2021aperiodic}. We found that the quality factors for the initial GA Vogel spiral and for the optimized photonic patch are $\mathrm{Q}_i\approx177$ and $\mathrm{Q}_o\approx384$, respectively. Figs. \ref{fig:LDOS_TE}(c) and \ref{fig:LDOS_TE}(d) show the spatial distributions of the Purcell factors (i.e., LDOS maps) of the initial and optimized structures, computed by using a square grid of excitation dipoles with a spacing of 3 nm, oriented in $\hat{\mathbf{x}}$ and $\hat{\mathbf{y}}$ directions \cite{dal2021aperiodic}. Figs. \ref{fig:LDOS_TE}(e) and \ref{fig:LDOS_TE}(f) show the spatial distributions of the electric fields of the optical resonances (normalized to their maximum values) corresponding to the initial and optimized structures, respectively. 

\begin{figure}[t!]
	\centering
	\includegraphics[width=\linewidth]{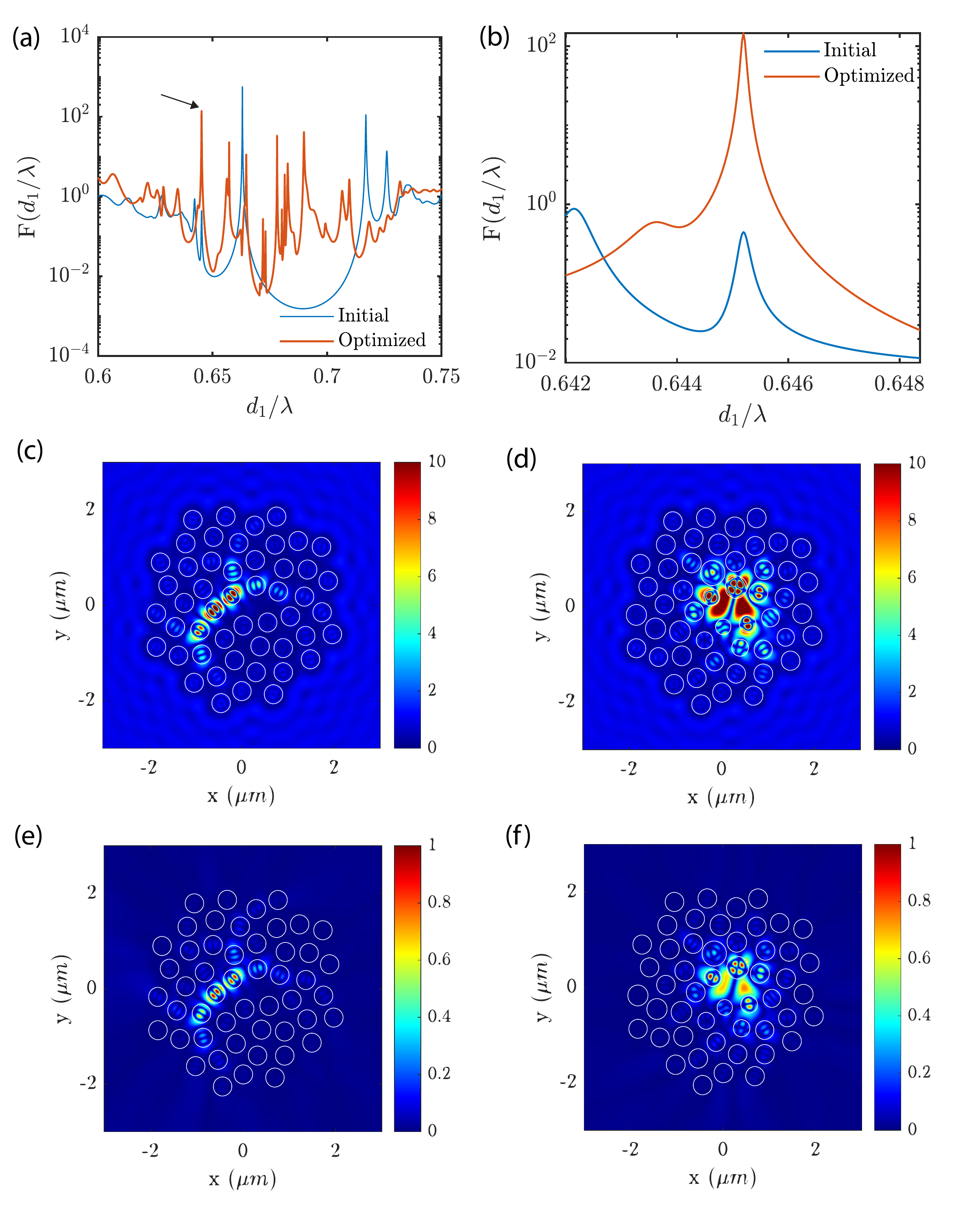}
	\caption{(a) Purcell factor spectrum for both the initial (blue) and optimized (red) GA Vogel spiral nanocylinder array with TM polarized dipole placed at $(0.0587\,\mu\mathrm{m}, 0.0352\,\mu\mathrm{m})$. The black arrow indicates the spectral parameter $d_1/\lambda=0.645193$ where we performed adjoint optimization. (b) Purcell factor spectrum for the initial (blue) and optimized (red) photonic patch near the optimized mode. The spatial map of Purcell factors of (c) initial and (d) optimized photonic patches at $d_1/\lambda=0.645193$. Also shown are the spatial distributions of TM-polarized optical modes at $d_1/\lambda=0.645193$ for the (e) initial and (f) optimized photonic patches, respectively.}
	\label{fig:LDOS_TM}
\end{figure}

\begin{figure}[t!]
	\centering
	\includegraphics[width=\linewidth]{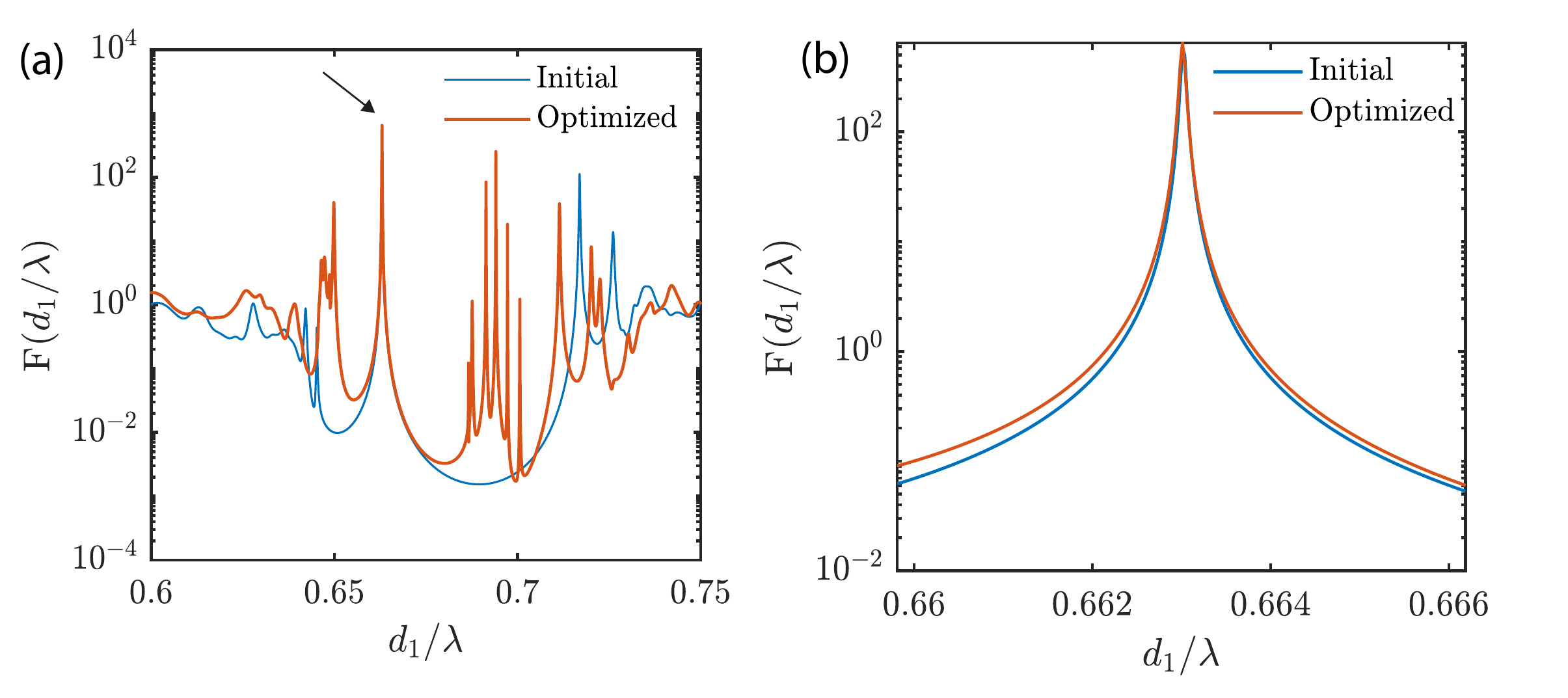}
	\caption{(a) Purcell factor spectrum for both the initial (blue) and optimized (red) GA Vogel spiral nanocylinder array with TM polarized dipole placed at $(0.0587\,\mu\mathrm{m}, 0.0352\,\mu\mathrm{m})$. The black arrow indicates the spectral parameter $d_1/\lambda=0.660311$ where we performed adjoint optimization. (b) Purcell factor spectrum for the initial (blue) and optimized (red) photonic patch near the targeted mode.}
	\label{fig:LDOS_TM_peak}
\end{figure}

In order to demonstrate the robustness of our design method we additionally present the optimization of TM-polarized modes in arrays of dielectric rods with large refractive index contrast \cite{oliver2020cavity, oliver2021cavity}. Specifically, as the initial configuration we choose the same GA Vogel spiral geometry discussed above but  considered 50 dielectric cylinders with large permittivity $\epsilon_n=12.8$ embedded in air. 
The spatial and spectral localization properties of the band-edge modes of GA Vogel spiral structures have been intensively investigated in nanophotonics as a viable approach to enable enhanced light-matter interactions over multiple-length scales  \cite{wang2018spectral, sgrignuoli2020multifractality, barber2008aperiodic, dal2013optics, prado2021structural}.
We chose the location of the excitation dipole at $(x_s,y_s)=(0.0587\,\mu\mathrm{m}, 0.0352\,\mu\mathrm{m})$ in order to evaluate the TM Purcell enhancement. The utilized learning rates and number of iterations were kept the same as in the previously discussed TE case. In Fig. \ref{fig:LDOS_TM}(a) we display in logarithmic scale the spectra of the Purcell factors for the initial and the optimized configurations at the spectral parameter $d_1/\lambda=0.645193$, indicated by the black arrow. In Fig. \ref{fig:LDOS_TM}(b) we compare the spectra of the Purcell factors of the initial and the optimal photonic patch configurations in a spectral region around the targeted mode. Our results show that the peak value for the initial array is $\mathrm{F}_i\approx0.447$ while the one of the optimized array is $\mathrm{F}_o\approx145$, demonstrating a $324\times$ enhancement. Moreover, we obtained $\mathrm{Q}_i\approx2944$ and $\mathrm{Q}_o\approx6435$ for the initial and for the optimized photonic patches, respectively. We also show the LDOS maps of the initial and optimized structures, excited by a grid of $\hat{\mathbf{z}}$-oriented dipoles with same spacing as the TE-polarized case, in Figs. \ref{fig:LDOS_TM}(c) and \ref{fig:LDOS_TM}(d). Finally, Figs. \ref{fig:LDOS_TM}(e) and \ref{fig:LDOS_TM}(f) display the spatial distributions of the resonant modes for the initial and optimized structures, respectively. For a more complete study of the LDOS enhancement, we also investigated the performance of the adjoint optimization on the most localized band-edge mode of the initial GA Vogel spiral. Fig. \ref{fig:LDOS_TM_peak}(a) shows the spectra of Purcell factors for the initial and the optimized structures at $d_1/\lambda=0.645193$ and we display the effect of the Purcell factor optimization over a smaller spectral region around the selected mode in Fig. \ref{fig:LDOS_TM_peak}(b). Our results show that the optimization improvement for this mode is very modest, with the Purcell factor increasing from $\mathrm{F}_i\approx576$ to $\mathrm{F}_o\approx647$. Therefore, we have established that Vogel spiral photonic patches already support a strongly localized band-edge mode with an almost optimal Purcell factor   $\mathrm{F}\approx\mathrm{Q}/V$, consistently with our previous studies \cite{trevino2012geometrical,liew2011localized,fab2019localization}.

\begin{figure}[t!]
	\centering
	\includegraphics[width=\linewidth]{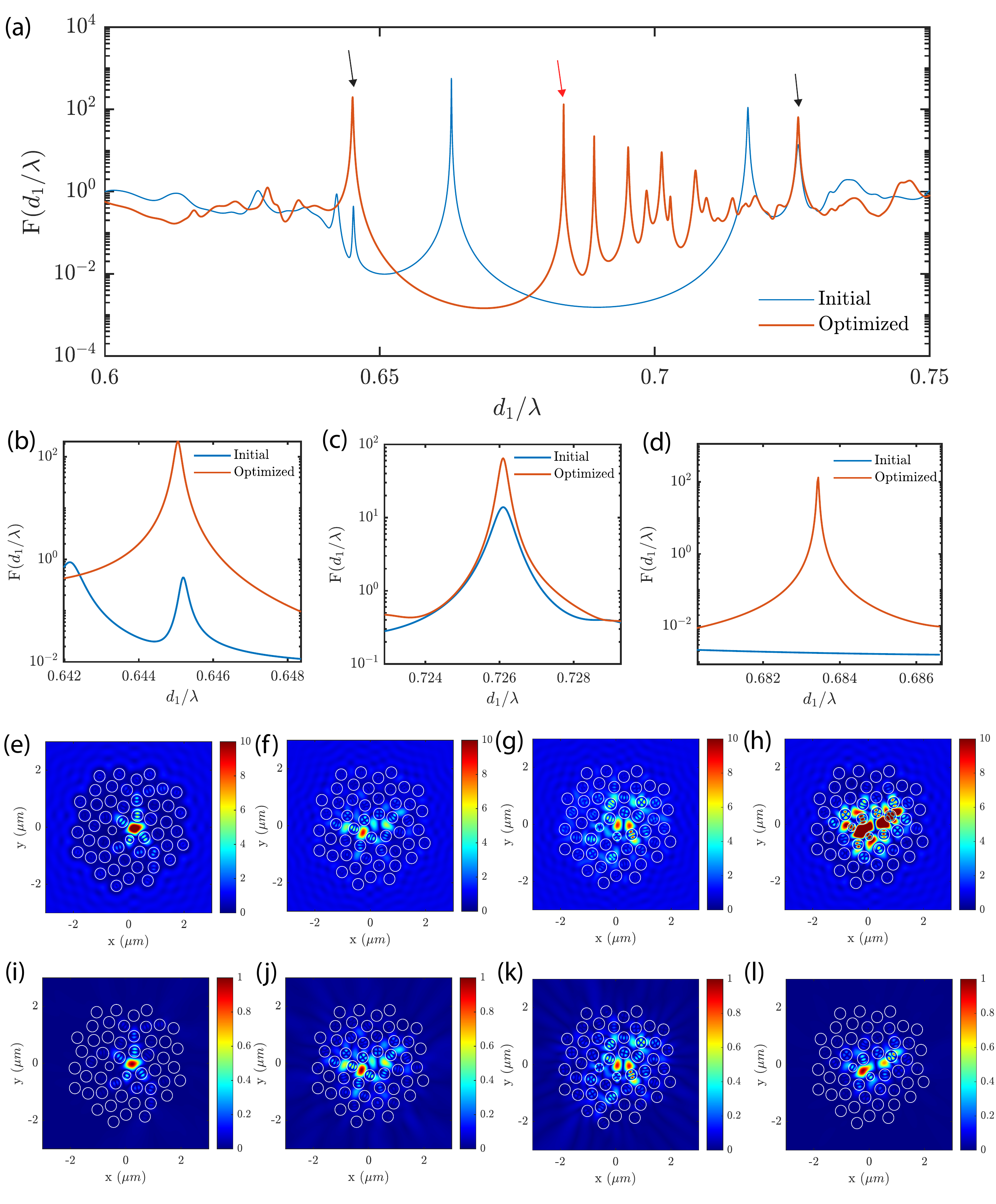}
	\caption{(a) Purcell factor spectrum for both the initial (blue) and optimized (red) GA Vogel spiral nanocylinder array with TM polarized dipole placed at $(0.0587\,\mu\mathrm{m}, 0.0352\,\mu\mathrm{m})$. The two black arrows indicate the specified spectral parameters $d_1/\lambda=0.645183, 0.726096$ where we performed adjoint optimization. Purcell factor spectrum for the initial (blue) and optimized (red) photonic patch in the spectral regions around (b) $d_1/\lambda=0.645183$, (c) $d_1/\lambda=0.726096$, and (d) $d_1/\lambda=0.683439$ (indicated by red arrow in (a)). LDOS maps of the (e) optimized structure at $d_1/\lambda=0.645183$, (f) initial and (g) optimized structures at $d_1/\lambda=0.726096$, and (h) optimized structure at $d_1/\lambda=0.683439$. (i)-(l) The spatial distributions of TM-polarized optical resonant modes, corresponding to the conditions in (e)-(h) respectively.}
	\label{fig:LDOS_TM_multi}
\end{figure}

In order to further explore the capabilities of our inverse design approach we investigate the possibility of enhancing the LDOS of photonic patches at multiple wavelengths. Similarly to the situation of broadband focusing discussed in subsection \ref{section: adjoint 2D-GMT}\ref{subsection: focusing photonic patch}, we introduce a  multi-objective function for the optimization of the Purcell factor at multiple wavelengths as follows:
\begin{equation}
    \mathrm{g}_\mathrm{TM}=\sum_{i=1}^{m}\mathrm{g}_\mathrm{TM}(\lambda_i)+\sum_{i\ne j}\left[\mathrm{F}(\textbf{r};\lambda_i)-\mathrm{F}(\textbf{r};\lambda_j)\right]^2
\end{equation}
where $m$ is the number of considered wavelengths and the cross difference penalty term is used to minimize the discrepancy between Purcell factors at different wavelengths. We chose the same initial array and dipole excitation conditions as in the optimization of the TM-polarized single mode discussed before. Additionally, all the learning rate parameters are identical to the previous case, but given the more challenging nature of this problem we ran the optimization algorithm for 70000 iterations. 
In in Fig. \ref{fig:LDOS_TM_multi}(a) we show the spectrum of the Purcell enhancement factor obtained when optimizing at the two spectral parameter values  $d_1/\lambda=0.645183, 0.726096$ indicated by the black arrows. The enhancement achieved at both the corresponding wavelengths is directly evident in Figs. \ref{fig:LDOS_TM_multi}(b) and \ref{fig:LDOS_TM_multi}(c). In particular, at $d_1/\lambda=0.645183$ the Purcell factor was increased from $\mathrm{F}_i\approx0.441$ to $\mathrm{F}_o\approx199$, which corresponds to a $451\times$ enhancement. The corresponding quality factor was improved from $\mathrm{Q}_i\approx2944$ to $\mathrm{Q}_o\approx3431$. On the other hand in \ref{fig:LDOS_TM_multi}(c), the Purcell factor at $d_1/\lambda=0.726096$ was increased from $\mathrm{F}_i\approx13.8$ to $\mathrm{F}_o\approx64.7$, achieving an overall LDOS enhancement of a factor of $4.68$. The corresponding quality factor was enhanced from $\mathrm{Q}_i\approx1250$ to $\mathrm{Q}_o\approx2689$. In Figs. \ref{fig:LDOS_TM_multi}(e) through \ref{fig:LDOS_TM_multi}(g) we display the spatial distributions of the LDOS maps at these two spectral parameters respectively, while in Figs. \ref{fig:LDOS_TM_multi}(i) through \ref{fig:LDOS_TM_multi}(k) we illustrate the corresponding optical modes. Note that a similar optimization can also be applied to the TE-polarized mode as well. 

It is worth noticing that after the optimization, resonant modes start to emerge inside the optical bandgap of the initial GA Vogel spiral structure. As an instance, we investigated a resonant mode located at $d_1/\lambda=0.683439$, which is indicated by the red arrow in Fig. \ref{fig:LDOS_TM_multi}(a). Fig. \ref{fig:LDOS_TM_multi}(d) indicates the optimzed Purcell factor with $\mathrm{F}_o\approx132$. The corresponding quality factor is $\mathrm{Q}_o\approx1.66\times10^4$. Fig. \ref{fig:LDOS_TM_multi}(h) and \ref{fig:LDOS_TM_multi}(l) show the LDOS maps and the resonant modes excited at $d_1/\lambda=0.683439$, respectively. 

\begin{figure}[t!]
	\centering
	\includegraphics[width=\linewidth]{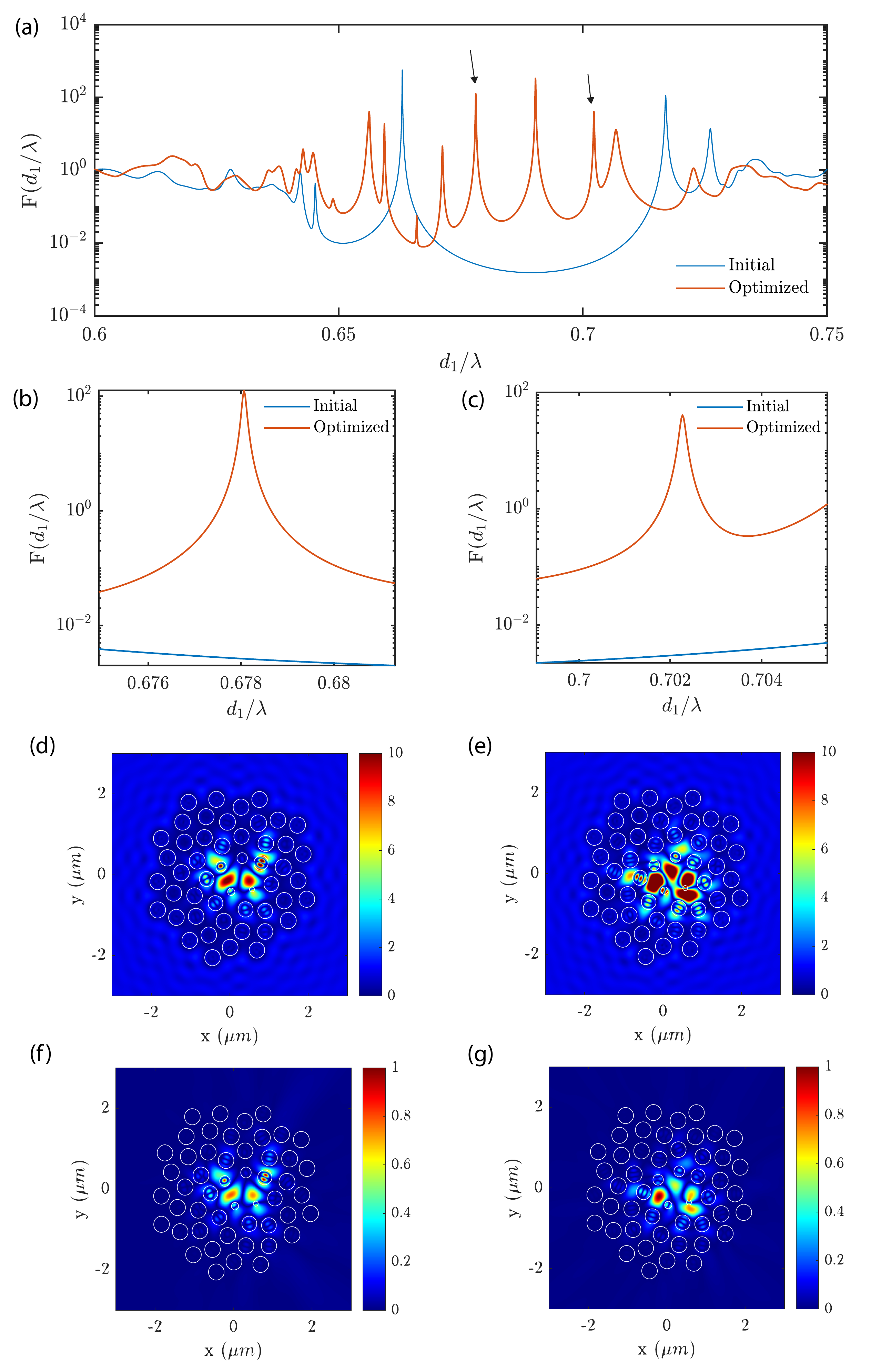}
	\caption{(a) Purcell factor spectrum for both the initial (blue) and optimized (red) GA Vogel spiral nanocylinder array with TM polarized dipole placed at $(0.0587\,\mu\mathrm{m}, 0.0352\,\mu\mathrm{m})$. The two black arrows indicate the specified spectral parameters $d_1/\lambda=0.678138, 0.702267$ where we performed adjoint optimization. Purcell factor spectrum for the initial (blue) and optimized (red) photonic patch in the spectral regions around (b) $d_1/\lambda=0.678138$ and (c) $d_1/\lambda=0.702267$. Optimized LDOS maps at (d) $d_1/\lambda=0.678138$ and (e) $d_1/\lambda=0.702267$. Also shown are the spatial distributions of TM-polarized optical modes at (f) $d_1/\lambda=0.678138$ and (g) $d_1/\lambda=0.702267$, respectively.}
	\label{fig:LDOS_TM_bandgap}
\end{figure}

Finally, in Fig. \ref{fig:LDOS_TM_bandgap}(a) we optimize the Purcell factor at two frequencies (i.e., at the corresponding spectral parameters indicated by the black arrows) that fall within the bandgap of the initial Vogel spiral structure.  Specifically, the selected spectral parameters are $d_1/\lambda=0.678138$ and $d_1/\lambda=0.702267$, and we considered 70000 iterations of our optimization algorithm. Figs. \ref{fig:LDOS_TM_bandgap}(b) and \ref{fig:LDOS_TM_bandgap}(c) show that optimized Purcell factors at $d_1/\lambda=0.645183$ and $d_1/\lambda=0.726096$ where we obtained $\mathrm{F}_o\approx126$ and $\mathrm{F}_o\approx40.6$, respectively. Their corresponding optimized quality factors are $\mathrm{Q}_o\approx6535$ to $\mathrm{Q}_o\approx4807$. The LDOS maps of the optimized structures at $d_1/\lambda=0.645183$ and $d_1/\lambda=0.726096$ are illustrated in Figs. \ref{fig:LDOS_TM_bandgap}(d) and \ref{fig:LDOS_TM_bandgap}(e), respectively. Figs. \ref{fig:LDOS_TM_bandgap}(f) and \ref{fig:LDOS_TM_bandgap}(g) display the selected optical modes for the optimized structures at the same two spectral parameters, respectively. 

\begin{figure}[t!]
	\centering
	\includegraphics[width=\linewidth]{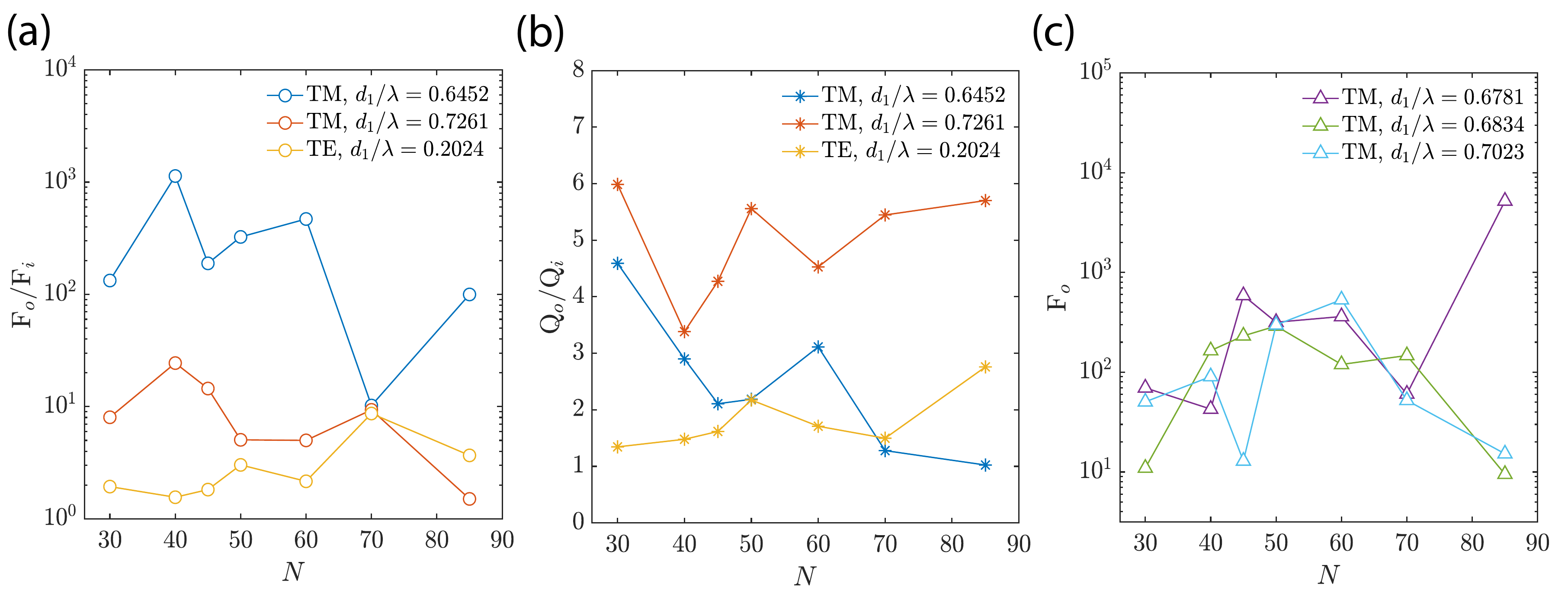}
	\caption{(a) Purcell factor enhancement $\mathrm{F}_o/\mathrm{F}_i$ in semilog scale and (b) quality factor enhancement $\mathrm{Q}_o/\mathrm{Q}_i$ with respect to number of scatterers $N$, for TM modes at $d_1/\lambda=0.645183, 0.726096$ and TE mode at $d_1/\lambda=0.202363$. (c) Purcell factor after optimization $\mathrm{F}_o$ in semilog scale with respect to number of scatterers $N$, for TM modes at $d_1/\lambda=0.678138, 0.683439,$ and $0.702267$. }
	\label{fig:LDOS_analysis_Nparticles}
\end{figure}

To conclude our study we analyze in Fig. \ref{fig:LDOS_analysis_Nparticles}(a) the scaling of the ratio between the optimized Purcell factor and initial Purcell factor $\mathrm{F}_o/\mathrm{F}_i$ as a function of the number of cylinders $N$ in the photonic patch. To perform this analysis, we considered the same modes and structure optimized in this subsection. Specifically, we investigated the optimization performance of TM modes at $d_1/\lambda=0.645193$ (as in Fig. \ref{fig:LDOS_TM}(b)), $d_1/\lambda=0.726096$ (as in Fig. \ref{fig:LDOS_TM_multi}(c)) and TE mode at $d_1/\lambda=0.202363$ (as in Fig. \ref{fig:LDOS_TE}(b)) for structures with different $N$. We remark that in these configurations the optimized modes are strongly localized in a small area at the center of the photonic patch. Consistently, due to the initial decrease of the optimized mode volume, we found that the ratio $\mathrm{F}_o/\mathrm{F}_i\approx (\mathrm{Q}_o V_i)/(\mathrm{Q}_i V_o)$ features a peak for photonic patches with an optimal size, which depends on the spectral parameter $d_1/\lambda$ of the mode. Beyond this point, the optimization enhancement decreases because the size of the photonic patch exceeds the characteristic localization length of the considered mode. In Fig. \ref{fig:LDOS_analysis_Nparticles}(b) we display the enhancement of the ratios of the corresponding quality factors $\mathrm{Q}_o/\mathrm{Q}_i$. In the case of optimized TM mode at $d_1/\lambda=0.645193$ (blue curve), we observe a decreasing trend with respect to the number of cylinders $N$. As $N$ increases, we found that $\mathrm{Q}_o/\mathrm{Q}_i$ decreases and it converges to unity when the photonic patches have $\approx 100$ cylinders. For the optimized TM mode at $d_1/\lambda=0.726096$ (red curve), the quality factor ratio decreases at first and then increases until it reaches a nearly constant value, as $N$ increases. As for the TE mode at $d_1/\lambda=0.202363$ (yellow curve), we notice that the ratio generally increases as $N$ increases. This behaviour can be explained by the fact that we are considering the properties of a resonant mode that is strongly localized in the central region of the photonic patch, as shown in Fig. \ref{fig:LDOS_TM}(d). In fact, this mode is characterized by a small localization length $\ell_{loc}$ of the order of only a few cylinders \cite{trevino2012geometrical, oliver2020cavity, dalnegro2022waves}. Therefore, depending on the value of $\ell_{loc}$, there exist a critical size for the photonic patch beyond which the benefits of mode optimization are essentially lost. Since $\mathrm{F}\approx \mathrm{Q}/V$, where $V$ is the mode volume, this occurs when the size of the device exceeds the characteristic localization length of the considered mode. Under these circumstances, the localized modes decouple from the rest of the structure, consistently with Fig. \ref{fig:LDOS_analysis_Nparticles}(a). On the other hand, we established that the most effective region of optimization with respect to $N$ varies from mode to mode, as shown in Fig. \ref{fig:LDOS_analysis_Nparticles}(b). Lastly, we present the scaling analysis of mode optimization inside the bandgap. Particularly, we investigated the TM modes at $d_1/\lambda=0.678138, 0.683439,$ and $0.702267$, which correspond to Fig. \ref{fig:LDOS_TM_bandgap}(b), Fig. \ref{fig:LDOS_TM_multi}(d) and Fig. \ref{fig:LDOS_TM_bandgap}(c).  Since there are no localized modes inside the bandgap of the initial GA Vogel spiral patches, their initial Purcell factors cannot be defined in this case. Hence, we show only the Purcell factors of the optimized structures in Fig. \ref{fig:LDOS_analysis_Nparticles}(c). We notice that as $\mathrm{F}_o$ changes with $N$, the curves feature optimal regions that are consistent with Fig. \ref{fig:LDOS_analysis_Nparticles}(a) with respect to the given sizes of photonic patches, depending on the chosen spectral parameters of the modes.

\section{Conclusions}
In this paper, we proposed and demonstrated a robust photonic inverse design method by combining adjoint optimization with rigorous semi-analytical 2D-GMT. We reviewed the GMT formalism in detail and derived closed-form analytical expressions that enable the efficient application of the gradient-based adjoint optimization of far-field and near-field relevant properties of photonic patches. We focused on multi-wavelength radiation shaping, near-field focusing, and the enhancement of the local density of states. 
Specifically, we demonstrated efficient far-field radiation shaping at multiple wavelengths in photonic patches optimized starting from both periodic and Vogel spiral configurations. Moreover, we designed compact focusing structures using both optimized aperiodic and periodic patches that enhance the field intensity at specified locations in the Fresnel zone with focusing efficiencies in excess of 75\%. We also demonstrated improved broadband focusing performances in optimized GA Vogel spiral patches. Lastly, we presented the design of optimized patches that enhance the LDOS and mode localization with both TE- and TM-polarized excitations at multiple wavelengths. We finally investigated the scaling of the optimized performances of photonic patches of different sizes. The combination of the semi-analytical 2D-GMT method with the adjoint optimization algorithm provides a robust inverse design methodology to develop compact photonic devices with optimal functionalities.
Without the need of spatial meshing, the developed approach provides efficient multiple scattering solutions with a strongly reduced computational burden compared to standard numerical simulation techniques and enables novel and more compact geometries for on-chip photonics and metamaterials device technologies.

\appendix
\section{Derivative of the transfer matrix}
\label{appendix: T derivative}
\begin{eqnarray}
    \frac{\partial \hat{\mathbf{T}}_{n n^{\prime}}^{\ell \ell^{\prime}}}{\partial p_{i}}=&&-\left(1-\delta_{n n^{\prime}}\right) e^{j\left(\ell^{\prime}-\ell\right) \phi_{n n^{\prime}}}\left[j\left(\ell^{\prime}-\ell\right) H_{\ell-\ell^{\prime}}\left(k_{o} R_{n n^{\prime}}\right) s_{n \ell}\right. \nonumber\\ && \times\frac{J_{\ell^{\prime}}\left(k_{o} r_{n^{\prime}}\right)}{J_{\ell}\left(k_{o} r_{n}\right)} \frac{\partial \phi_{n n^{\prime}}}{\partial p_{i}}
    +k_{o} H_{\ell-\ell^{\prime}}^{\prime}\left(k_{o} R_{n n^{\prime}}\right) s_{n \ell} \frac{J_{\ell^{\prime}}\left(k_{o} r_{n^{\prime}}\right)}{J_{\ell}\left(k_{o} r_{n}\right)} \nonumber\\
    && \times \frac{\partial R_{n n^{\prime}}}{\partial p_{i}} +H_{\ell-\ell^{\prime}}\left(k_{o} R_{n n^{\prime}}\right) \frac{J_{\ell^{\prime}}\left(k_{o} r_{n^{\prime}}\right)}{J_{\ell}\left(k_{o} r_{n}\right)} \frac{\partial s_{n \ell}}{\partial p_{i}} \nonumber\\
    &&+k_{o} H_{\ell-\ell^{\prime}}\left(k_{o} R_{n n^{\prime}}\right) s_{n \ell} \frac{J_{\ell^{\prime}}^{\prime}\left(k_{o} r_{n^{\prime}}\right)}{J_{\ell}\left(k_{o} r_{n}\right)} \frac{\partial r_{n^{\prime}}}{\partial p_{i}} \nonumber\\
    &&-k_{o} H_{\ell-\ell^{\prime}}\left(k_{o} R_{n n^{\prime}}\right) s_{n \ell} \frac{J_{\ell^{\prime}}\left(k_{o} r_{n^{\prime}}\right)}{J_{\ell}\left(k_{o} r_{n}\right)^{2}} J_{\ell}^{\prime}\left(k_{o} r_{n}\right) \frac{\partial r_{n}}{\partial p_{i}}
\end{eqnarray}
where the derivatives of $s_{n\ell}$ and $\Gamma_{n\ell}$ are as follows:
\begin{eqnarray}
    \frac{\partial s_{n \ell}}{\partial p_{i}}=&&-\left[k_{o} J_{\ell}^{\prime \prime}\left(k_{o} r_{n}\right) \frac{\partial r_{n}}{\partial p_{i}}-\frac{\partial \Gamma_{n \ell}}{\partial p_{i}} J_{\ell}\left(k_{o} r_{n}\right)\right.\nonumber\\
    &&\left.-k_{o} \Gamma_{n \ell} \frac{\partial r_{n}}{\partial p_{i}} J_{\ell}^{\prime}\left(k_{o} r_{n}\right)\right]\bigg/\left[{H_{\ell}^{\prime}\left(k_{o} r_{n}\right)-\Gamma_{n \ell} H_{\ell}\left(k_{o} r_{n}\right)}\right] \nonumber\\
    && +{\left[k_{o} H_{\ell}^{\prime \prime}\left(k_{o} r_{n}\right) \frac{\partial r_{n}}{\partial p_{i}}-\frac{\partial \Gamma_{n \ell}}{\partial p_{i}} H_{\ell}\left(k_{o} r_{n}\right)-k_{o} \Gamma_{n \ell} \frac{\partial r_{n}}{\partial p_{i}} \right.}\nonumber\\ &&\times H_{\ell}^{\prime}\left(k_{o} r_{n}\right)\bigg] \bigg/{\frac{\left(H_{\ell}^{\prime}\left(k_{o} r_{n}\right)-\Gamma_{n \ell} H_{\ell}\left(k_{o} r_{n}\right)\right)^{2}}{J_{\ell}^{\prime}\left(k_{o} r_{n}\right)-\Gamma_{n \ell} J_{\ell}\left(k_{o} r_{n}\right)}} 
    \label{eq:snlDerivative}\\
    \frac{\partial \Gamma_{n \ell}}{\partial p_{i}} =  &&\xi_{n} \frac{k_{n}^{2}}{k_{o}} \frac{\partial r_{n}}{\partial p_{i}}\left[\frac{J_{\ell}^{\prime \prime}\left(k_{n} r_{n}\right)}{J_{\ell}\left(k_{n} r_{n}\right)}-\frac{J_{\ell}^{\prime}\left(k_{n} r_{n}\right)^{2}}{J_{\ell}\left(k_{n} r_{n}\right)^{2}}\right]
\end{eqnarray}
The derivatives of $R_{nn'},\phi_{nn'}$ with respect to the different geometrical parameters can be found in Table \ref{table:derivativesCylrRelPos} below:

\begin{table}[h!]
    \caption{\label{table:derivativesCylrRelPos}%
        Derivatives of relative cylinder positions with respect to design parameters.}
    \begin{tabular}{ll}
        \hline
        $\begin{aligned}\\ 
            \frac{\partial R_{nn'}}{\partial x_j} &= \frac{x_n-x_{n'}}{R_{nn'}}(\delta_{jn}+\delta_{jn'})\\
            \frac{\partial R_{nn'}}{\partial y_j} &=\frac{y_n-y_{n'}}{R_{nn'}}(\delta_{jn}+\delta_{jn'})\\
            \frac{\partial R_{nn'}}{\partial r_j} &=0\\
        \end{aligned}$ &
        $\begin{aligned}\\
            \frac{\partial \phi_{nn'}}{\partial x_j} &=  \frac{\sin(\phi_{nn'})}{R_{nn'}}(\delta_{jn'}-\delta_{jn})\\
            \frac{\partial \phi_{nn'}}{\partial y_j} &=  \frac{\cos(\phi_{nn'})}{R_{nn'}}(\delta_{jn'}-\delta_{jn})\\
            \frac{\partial \phi_{nn'}}{\partial r_j} &=0\\
        \end{aligned}$\\
        \hline
    \end{tabular}
\end{table}

\section{Derivative of the plane wave excitation coefficients}
\label{appendix: plane wave derivative}
For a plane wave propagating at an angle $\Theta$ with respect to the $+\hat{x}$ in the 2D geometry, the coefficient $\hat{\mathbf{a}}^{0}$ is given by \cite{gagnon_JO_2015}:
\begin{equation}
    \hat{a}^{0E}_{n\ell}=\frac{a_{n \ell}^{0 E}}{J_{\ell}\left(k_{o} r_{n}\right)}=\frac{j^le^{j\mathbf{k}_o\cdot\mathbf{R}_n}e^{-j\ell \Theta}}{J_{\ell}\left(k_{o} r_{n}\right)}
\end{equation}
where $\mathbf{k}_o=k_o \cos(\Theta)\hat{\mathbf{x}}+k_o\sin(\Theta)\hat{\mathbf{y}}$ is the wavenumber in the host medium and $\mathbf{R}_n=(x_n,y_n)$ is the position of the $n\mathrm{th}$ cylinder.
The derivatives of the source expansion coefficients are given by:

\begin{eqnarray}
    \frac{\partial \hat{\mathbf{a}}^{0}}{\partial p_{i}} =\frac{a_{n \ell}^{0 E}}{J_{\ell}\left(k_{o} r_{n}\right)}\left[j s_{n \ell} \mathbf{k}_{o} \cdot \frac{\partial \mathbf{R}_{n}}{\partial p_{i}} + \frac{\partial s_{n \ell}}{\partial p_{i}}\nonumber \right.\\
    \left.-s_{n \ell} k_{o} \frac{J_{\ell}^{\prime}\left(k_{o} r_{n}\right)}{J_{\ell}\left(k_{o} r_{n}\right)} \frac{\partial r_{n}}{\partial p_{i}}\right]
\end{eqnarray}
where the partial derivative of $s_{n\ell}$ is given in Eq. \ref{eq:snlDerivative} and the partial derivatives of $r_{n}$ are given in the Table \ref{table:derivativesCylGeom}.

\section{Derivative of the dipole excitation coefficients}
\label{appendix: dipole derivative}
The source coefficients $\mathbf{\hat{a}}^0$ for a dipole in the host medium with different orientations are as follows \cite{asatryan2003}:
\begin{eqnarray}
    a_{n \ell, x}^{0 E} =&& -\frac{1}{8 j}\left[H_{\ell+1}\left(k_{o} R_{n s}\right) e^{-j(\ell+1) \theta_{n s}} \right.\nonumber
    \\ &&\left.+H_{\ell-1}\left(k_{o} R_{n s}\right) e^{-j(\ell-1) \theta_{n s}}\right] \\
    a_{n \ell, y}^{0 E} =&&-\frac{1}{8}\left[H_{\ell+1}\left(k_{o} R_{n s}\right) e^{-j(\ell+1) \theta_{n s}} \right.\nonumber\\
    && \left.-H_{\ell-1}\left(k_{o} R_{n s}\right) e^{-j(\ell-1) \theta_{n s}}\right] \\
    a_{n \ell, z}^{0 E} =&&\frac{1}{4 j} H_{\ell}\left(k_{o} R_{n s}\right) e^{-j \ell \theta_{n s}}
\end{eqnarray}
where $(R_{is},\theta_{is}$) are the polar coordinates of the source position ($x_s,y_s$) in the frame of reference of the $i\mathrm{th}$ cylinder center. The subscripts $x,y,z$ indicates the dipole orientation. 

Therefore, the derivatives of $\hat{\mathbf{a}}^{0}$ for exterior dipole sources along different orientations are computed as follows:

\begin{eqnarray}
    &&\frac{\partial \hat{\mathbf{a}}_{x}^{0}}{\partial p_{i}}=\hat{a}_{n \ell, x}^{0 E}\left[\frac{1}{s_{n \ell}} \frac{\partial s_{n \ell}}{\partial p_{i}}-k_{o} \frac{J_{\ell}^{\prime}\left(k_{o} r_{n}\right)}{J_{\ell}\left(k_{o} r_{n}\right)} \frac{\partial r_{n}}{\partial p_{i}}\right]-\frac{1}{8 j} e^{-j(\ell+1)\theta_{n s}} \nonumber\\
    && \times \left(k_{o} H_{\ell+1}^{\prime}\left(k_{o} R_{n s}\right) \frac{\partial R_{n s}}{\partial p_{i}}-j(\ell+1) H_{\ell+1}\left(k_{o} R_{n s}\right) \frac{\partial \theta_{n s}}{\partial p_{i}}\right)\nonumber\\    
    &&-\frac{1}{8 j} e^{-j(\ell-1) \theta_{n s}}\left(k_{o} H_{\ell-1}^{\prime}\left(k_{o} R_{n s}\right) \frac{\partial R_{n s}}{\partial p_{i}}-j(\ell-1) \right.\nonumber
    \\
    &&\left.\times H_{\ell-1}\left(k_{o} R_{n s}\right) \frac{\partial \theta_{n s}}{\partial p_{i}}\right) \\
    &&\frac{\partial \hat{\mathbf{a}}_{y}^{0}}{\partial p_{i}}=\hat{a}_{n \ell, y}^{0 E}\left[\frac{1}{s_{n \ell}} \frac{\partial s_{n \ell}}{\partial p_{i}}-k_{o} \frac{J_{\ell}^{\prime}\left(k_{o} r_{n}\right)}{J_{\ell}\left(k_{o} r_{n}\right)} \frac{\partial r_{n}}{\partial p_{i}}\right] -\frac{1}{8} e^{-j(\ell+1) \theta_{n s}}\nonumber\\
    &&\times\left(k_{o} H_{\ell+1}^{\prime}\left(k_{o} R_{n s}\right) \frac{\partial R_{n s}}{\partial p_{i}}-j(\ell+1) H_{\ell+1}\left(k_{o} R_{n s}\right) \frac{\partial \theta_{n s}}{\partial p_{i}}\right) \nonumber\\
    &&+\frac{1}{8} e^{-j(\ell-1) \theta_{n s}}\left(k_{o} H_{\ell-1}^{\prime}\left(k_{o} R_{n s}\right) \frac{\partial R_{n s}}{\partial p_{i}}-j(\ell-1) \nonumber\right.\\
    && \left.H_{\ell-1}\left(k_{o} R_{n s}\right) \frac{\partial \theta_{n s}}{\partial p_{i}}\right)\\
    &&\frac{\partial \hat{\mathbf{a}}_{z}^{0}}{\partial p_{i}}=\hat{a}_{n \ell, z}^{0 E}\left[\frac{1}{s_{n \ell}} \frac{\partial s_{n \ell}}{\partial p_{i}}-k_{o} \frac{J_{\ell}^{\prime}\left(k_{o} r_{n}\right)}{J_{\ell}\left(k_{o} r_{n}\right)} \frac{\partial r_{n}}{\partial p_{i}}\right] +\frac{1}{4 j} e^{-j \ell \theta_{n s}}\nonumber\\
    &&\times \left[k_{o} H_{\ell}^{\prime}\left(k_{o} R_{n s}\right) \frac{\partial R_{n s}}{\partial p_{i}}-j \ell H_{\ell}\left(k_{o} R_{n s}\right) \frac{\partial \theta_{n s}}{\partial p_{i}}\right]
\end{eqnarray}
where the derivative of $R_{n s}, \theta_{ns}$ with respect to the design parameters can be found in the Table \ref{table:derivativesDipole}.

\begin{table}[h!]
    \caption{\label{table:derivativesDipole}%
        Derivatives of relative cylinder positions with respect to design parameters.}
    \begin{tabular}{ll}
        \hline
        ${\begin{aligned} \\ \frac{\partial R_{n s}}{\partial x_{i}} &=-\cos \left(\theta_{n s}\right) \delta_{i n} & \frac{\partial \theta_{n s}}{\partial x_{i}} &=+\frac{\sin \left(\theta_{n s}\right)}{R_{n s}} \delta_{i n} \\ \frac{\partial R_{n s}}{\partial y_{i}} &=-\sin \left(\theta_{n s}\right) \delta_{i n} & \frac{\partial \theta_{n s}}{\partial y_{i}} &=-\frac{\cos \left(\theta_{n s}\right)}{R_{n s}} \delta_{i n} \\ \frac{\partial R_{n s}}{\partial r_{i}} &=0 & \frac{\partial \theta_{n s}}{\partial r_{i}} &=0 \end{aligned}}$\\
        \hline
    \end{tabular}
\end{table}
\begin{backmatter}
    \bmsection{Funding} L.D.N. acknowledges the support from the National Science Foundation (ECCS-2110204 and ECCS-2015700) and the U.S. Army Research Laboratory under Cooperative Agreement Number W911NF-12-2-0023.
    \bmsection{Acknowledgments} We wish to acknowledge the support of the author community in using REV\TeX{}, offering suggestions and encouragement and testing new versions. 
    \bmsection{Disclosures} The authors declare no conflicts of interest. 
    \bmsection{Data Availability Statement} The data that support the findings of this study are available from the corresponding author upon reasonable request. 
    
    
\end{backmatter}

\bibliography{apssamp}

\bibliographyfullrefs{apssamp}
\end{document}